\documentclass{article}
\usepackage{graphicx}
\usepackage{subfigure} 
\usepackage{natbib}
\usepackage{algorithm}
\usepackage{algorithmic}
\usepackage{hyperref}

\usepackage[accepted]{icml2012}

\usepackage{amssymb,amsmath}
\usepackage{pgfplots}
\pgfplotsset{compat=newest}
\usepgfplotslibrary{groupplots}
\usepackage{pgfplotstable}
\usepackage{booktabs}
\usepackage{colortbl}
\usepackage{verbatim}
\usepackage{multirow}
\usepackage{picinpar}
\usepackage{fmtcount}
\usepackage{float}

\makeatletter

\newcommand{\Rmnum}[1]{\expandafter\@slowromancap\romannumeral #1@}
\makeatother

\providecommand{\abs}[1]{\lvert#1\rvert}
\providecommand{\norm}[1]{\lVert#1\rVert}
\newcommand{\vc}[1]{{\pmb{#1}}}
\newcommand{\sign}{\operatorname{sign}}

\ifx\BlackBox\undefined
\newcommand{\BlackBox}{\rule{1.5ex}{1.5ex}}
\fi

\ifx\proof\undefined
\newenvironment{proof}{\par\noindent{\bf Proof\ }}{\hfill\BlackBox\\[2mm]}
\fi

\ifx\theorem\undefined
\newtheorem{theorem}{Theorem}
\fi

\ifx\lemma\undefined
\newtheorem{lemma}[theorem]{Lemma}
\fi

\ifx\definition\undefined
\newtheorem{definition}[theorem]{Definition}
\fi

\icmltitlerunning{Robust Classification with Adiabatic Quantum Optimization}

\begin{document} 

\twocolumn[
\icmltitle{Robust Classification with Adiabatic Quantum Optimization}

\icmlauthor{Vasil S.~Denchev}{denchev@gmail.com}
\icmladdress{Purdue University/Google}
\icmlauthor{Nan Ding}{ding10@purdue.edu}
\icmladdress{Purdue University}
\icmlauthor{S. V. N. Vishwanathan}{vishy@stat.purdue.edu}
\icmladdress{Purdue University}
\icmlauthor{Hartmut Neven}{neven@google.com}
\icmladdress{Google}

\icmlkeywords{supervised learning, robust classification, non-convex optimization, quantum adiabatic}

\vskip 0.3in
]

\begin{abstract}
We propose a non-convex training objective for robust binary classification of data sets in which label noise is present. The design is guided by the intention of solving the resulting problem by adiabatic quantum optimization. Two requirements are imposed by the engineering constraints of existing quantum hardware: training problems are formulated as quadratic unconstrained binary optimization; and model parameters are represented as binary expansions of low bit-depth. In the present work we validate this approach by using a heuristic classical solver as a stand-in for quantum hardware. Testing on several popular data sets and comparing with a number of existing losses we find substantial advantages in robustness as measured by test error under increasing label noise. Robustness is enabled by the non-convexity of our hardware-compatible loss function, which we name \emph{$q$-loss}. 
\end{abstract}

\section{Introduction}\label{introduction}
In recent years machine learning researchers and practitioners have been focusing on convex optimization methods due to their computational advantages and well understood mathematical properties. The many successes of convexity-based algorithms are witnesses to that. While it is easily recognized that allowing for non-convex objectives opens up a plethora of possibilities for better solutions to machine learning problems, much of the contemporary research has deliberately avoided them. The reason is the widely known fact that non-convexity often results in NP-hard problems. 

However this choice comes at a cost as the shortcomings of convex objectives are also well understood. Recent work \cite{LongServedio} showed that convex loss functions cannot be made robust in the presence of label noise because they cause unbounded growth of penalties for large negative margins. \cite{ManwaniSastry} further characterized this effect by analyzing various convex losses and found that none of them is tolerant to non-uniform label noise. In practice label noise turns out to be a serious problem due to the fact that it affects real-world data sets to a significant degree. Since label noise manifests itself throughout the optimization as large negative margins, the finally constructed decision hyperplane that represents the global minimum of any convex loss tends to be pulled by the mislabeled training examples away from the minimizer of classification error. Therefore, even though solving convex losses to optimality is feasible, when label noise causes the lowest objective value to not correspond to the lowest attainable training error, the entire exercise misses the mark. Consequently, any approach exhibiting this problem does not stand to benefit from improved optimization techniques.

\begin{figure}[t]
\centering
\tikzset{every mark/.append style={scale=0.5}}
\pgfplotsset{every axis/.append style={height=2.7cm,width=3.3cm}}
\begin{tikzpicture}
   \begin{axis} [name=plot1,tick label style={font=\tiny},xtick = {3.61,3.62},ytick={5.95,6.03,6.1}]
    \addplot coordinates {(3.605463,6.095900) (3.605463,6.095900)
    (3.605463,6.095900) (3.605463,6.095900) (3.605463,6.095900) (3.605463,6.095900) (3.605463,6.095900) (3.605463,6.095900) (3.605463,6.095900) (3.621708,6.071400) (3.622302,6.046900) (3.622481,6.022300) (3.622482,6.022300) (3.622482,6.022300) (3.622587,6.034600) (3.624878,5.948700) (3.624878,5.948700) (3.624878,5.936500) (3.624878,5.924200)};
  \end{axis}
  \begin{axis} [name=plot2,tick label style={font=\tiny},at={($(plot1.east)+(0.8cm,0)$)},anchor=west,xtick = {3.61,3.62}]
    \addplot coordinates {(3.607717,6.095900) (3.607717,6.095900) (3.607717,6.095900) (3.607717,6.095900) (3.607717,6.095900) (3.607717,6.095900) (3.607717,6.095900) (3.607717,6.095900) (3.617321,6.010100) (3.617685,6.046900) (3.618311,5.997800) (3.618514,5.973300) (3.622547,5.924200) (3.622547,5.924200) (3.622547,5.924200) (3.622549,5.899700) (3.622549,5.899700) (3.622550,5.887400) (3.622550,5.899700)};
  \end{axis}
  \begin{axis} [name=plot3,tick label style={font=\tiny},at={($(plot2.east)+(0.8cm,0)$)},anchor=west]
    \addplot coordinates {(3.512335,4.023100) (3.512335,4.023100) (3.512335,4.023100) (3.512335,4.023100) (3.512335,4.023100) (3.512335,4.023100) (3.512336,4.047600) (3.512336,4.047600) (3.512336,4.047600) (3.512336,4.047600) (3.531346,3.532400) (3.532142,3.507900) (3.532143,3.495600) (3.532143,3.495600) (3.532693,3.507900) (3.532693,3.507900) (3.534466,3.532400) (3.535221,3.520200) (3.535221,3.520200)};
    \end{axis}
  \begin{axis} [name=plot4,tick label style={font=\tiny},at={($(plot1.south)-(0,0.5cm)$)},anchor=north,xtick = {3.615,3.625},xlabel=1/Risk,ylabel style={xshift=22pt},ylabel=Training error ($\%$)]
    \addplot coordinates {(3.614818,5.985500) (3.614818,5.985500) (3.614818,5.985500) (3.614818,5.985500) (3.614818,5.985500) (3.614818,5.985500) (3.614818,5.985500) (3.614818,5.985500) (3.614996,5.973300) (3.623098,5.961000) (3.623182,5.973300) (3.623685,5.973300) (3.623685,5.973300) (3.623871,5.924200) (3.623989,5.924200) (3.627479,5.789300) (3.627572,5.826100) (3.627572,5.826100) (3.627577,5.826100)};   
     \end{axis}
  \begin{axis} [name=plot5,tick label style={font=\tiny},at={($(plot2.south)-(0,0.5cm)$)},anchor=north,ytick={4.7,5,5.4},xlabel=1/Risk]
    \addplot coordinates {(2.076761,5.384500) (2.076761,5.384500) (2.076761,5.384500) (2.076761,5.384500) (2.076761,5.384500) (2.076761,5.384500) (2.076761,5.384500) (2.076761,5.384500) (2.076761,5.384500) (2.111767,4.746700) (2.111965,4.759000) (2.114714,4.820300) (2.118439,4.943000) (2.118439,4.943000) (2.120929,5.077900) (2.123235,5.188300) (2.123235,5.188300) (2.124805,5.188300) (2.125519,5.200500)};
  \end{axis}
  \begin{axis} [name=plot6,tick label style={font=\tiny},at={($(plot3.south)-(0,0.5cm)$)},anchor=north,xtick = {3.04,3.08,3.12},xlabel=1/Risk,ytick={7.6,8,8.4}]
    \addplot coordinates {(3.035601,7.567800) (3.035601,7.567800) (3.035601,7.567800) (3.035601,7.567800) (3.035601,7.567800) (3.036419,7.592300) (3.036419,7.592300) (3.036419,7.592300) (3.036419,7.592300) (3.102356,7.911200) (3.114548,8.303700) (3.114548,8.303700) (3.114548,8.303700) (3.114548,8.303700) (3.114548,8.303700) (3.114548,8.303700) (3.115177,8.352800) (3.115177,8.352800) (3.115177,8.352800)};
   \end{axis}
\end{tikzpicture}
 \caption{Relationship between training error and inverse empirical risk produced by minimizing square loss on six different binary classifiers for digits (e.g. '1' vs the rest, '2' vs the rest, etc.) The data ("OCR in photos"; 10200 dimensions; 38924 examples; 10 classes) represents a challenging real-world training problem of significant practical importance. An adequate loss function should generally be decreasing training error as the empirical risk approaches global minimum (top plots). Unfortunately, the opposite effect (bottom plots) can often be observed when working with convex losses. The failures are found to be due to two factors, both of which cause square loss to be drastically misled by its convexity: occasionally mistaken labels resulting from the semi-automatic process generating the data; and the presence of examples of one class that may be similar to examples of another class (e.g. '6' and '8').}
\label{fig:non-monotonic}
\end{figure}
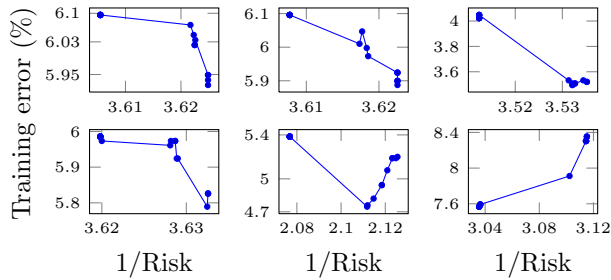

For example, Fig.~\ref{fig:non-monotonic} shows the broken correspondence between training error and objective value when a convex loss is used in a training problem of practical significance---"OCR in photos". The human task of tagging characters in photos of potentially poor quality is not easy, so the presence of mislabeled examples in the training set is not surprising. Even worse, routinely used semi-automatic preparation of training data is also contributing to mistakes. The problem may gradually disappear for cleaner data sets, which often happen to be the cases when convex losses produce excellent classifiers. Unfortunately the nature of large-scale supervised learning does not permit elaborate quality assurance for data sets that are handed out to training algorithms; accordingly label noise will continue to pollute real-world data sets. Moreover future intelligent systems will rely increasingly on weakly labeled or unlabeled data increasing the need for noise tolerance.

\cite{DingVishy}  and \cite{Hamed} took these lessons and independently studied two different non-convex but seemingly well behaved types of loss functions. \cite{Bottou2006, Bottou2011} also explored non-convexity in the context of SVM with ramp loss, but their focus was on achieving sparser sets of support vectors and speed of training rather than improved accuracy and robustness of the constructed classifier. 

In the present work we continue the study of non-convexity. We report on training with a non-convex objective using discrete optimization in a formulation adapted to take advantage of emerging hardware that performs adiabatic quantum optimization (AQO). AQO, first proposed in \cite{Farhi2000}, is a quantum computing model with good prospects for scalable and practically useful hardware implementation. Studies of its purported computational superiority over classical computing have repeatedly given encouraging results, e.g. \cite{AminDickson}. Significant investments are underway by the Canadian company D-Wave to develop a hardware implementation. A series of rigorous studies of the quantum mechanical properties of the D-Wave processors, culminating in a recent Nature publication \cite{dwave}, have increased the excitement in the quantum computing community for this approach. This was further fueled by news of a successful collaboration with Google \cite{CarDetector} and of Lockheed Martin purchasing an adiabatic quantum computer. For machine learning purposes, D-Wave's implementation of AQO can be regarded as a black-box discrete optimization engine that accepts any problems formulated as quadratic unconstrained binary optimization (QUBO), also equivalent to the Ising model and Weighted MAX-2-SAT. It should be noted that this training formulation is a good format for AQO independently of D-Wave's efforts since it can be physically realized as the simplest possible multi-qubit configuration---an Ising system \cite{Brush}. We do not claim principled superiority of q-loss over other non-convex losses. However $q$-loss is distinguished by the fact that it can be formulated for AQO on quantum hardware that only supports quadratic (2-local) interactions among its qubits using a number of ancillary variables that just grows linearly with the number of training examples. To the best of our current knowledge, no other non-convex loss has this property\footnote{Except for the non-margin-enforcing 0-1 loss \cite{Neven09}}. While all other non-convex losses are tackled by heuristic optimization with very limited success, $q$-loss may be solvable to optimality by AQO.

The paper is organized as follows: Section~\ref{training} defines the training problem; Section~\ref{q_loss_section} introduces $q$-loss, derives its QUBO formulation, and discusses the intuition behind it; Sections~\ref{bounding_q}~and~\ref{discrete_weights} deal with choosing hyper-parameter values and discretization of variables; Section~\ref{experiments} presents our experiments; and Section~\ref{conclusion} concludes with an overview and discussion. Technical details can be found in the supplementary material.

\section{Training a binary classifier}\label{training}
We study binary classifiers $y = \sign \left(\vc{w}^T\vc{x} + b\right)$, where $\vc{x} \in \mathbb{R}^N$ is an input pattern to be classified, $y \in \{-1,1\}$ is the label associated with $\vc{x}$, $\vc{w} \in \mathbb{R}^N$ is a vector of weights to be optimized, and $b \in \mathbb{R}$ is the bias. Training, also known as regularized risk minimization, consists of choosing $\vc{w}$ and $b$ by simultaneously minimizing two terms: \emph{empirical risk} $R(\vc{w}, b) = \sum_{s=1}^S L\left(m\left(\vc{x}_s, y_s, \vc{w}, b\right)\right)/S$ and \emph{regularization} $\Omega(\vc{w})$. $R$, via a \emph{loss function} $L$, estimates the error that any candidate classifier causes over a set of $S$ training examples $\{(\vc{x}_s, y_s) | s = 1, \ldots, S\}$. The argument of $L$ is known as the \emph{margin} of example $s$ with respect to the decision hyperplane defined by $\vc{w}$ and $b$:
\begin{equation}\label{margin}
m\left(\vc{x}_s, y_s, \vc{w}, b\right) = y_s \left(\vc{w}^T\vc{x}_s + b\right)
\end{equation}
$\Omega$ controls the complexity of the classifier and is necessary for good generalization because classifiers with high complexity display overfitting---they can classify the training set with low error but may not do well on previously unseen data. Training amounts to solving
\begin{equation}\label{train_opt}
(\vc{w}, b)^{*} = \arg\min_{\vc{w}, b} \left\{R\left(\vc{w}, b\right)+\Omega\left(\vc{w}\right) \right\} \hbox{ . }
\end{equation}

The most natural choice for $L$ is  \emph{0-1 loss}, which simply indicates a misclassification for a negative margin:
\begin{equation}\label{0_1_loss}
L_{\text{0-1}}(m) = \left(1 - \sign\left(m\right)\right)/2
\end{equation}
Due to the non-convexity of $L_{\text{0-1}}$, the resulting optimization problem \eqref{train_opt} is NP-hard \cite{Feldman}. To avoid dealing with NP-hard optimization problems, in practice $L_{\text{0-1}}$ is replaced by some convex upper bound (e.g. square, logistic, exponential, hinge), and $\Omega$ is usually chosen as $\ell_{1}$- or $\ell_{2}$-norm penalization of $\vc{w}$. This allows arriving at convex optimization problems that can be rigorously analyzed and efficiently solved by classical means. However, such relaxations are known to compromise the original goal of training because convex losses can be severely misled by label noise in the training data.

\section{$q$-loss}\label{q_loss_section}
Because the quantum hardware natively represents a general family of quadratic functions, the simplest loss function that would work is \emph{square loss}, which is a convex upper bound to $L_{\text{0-1}}$:
\begin{equation}\label{square_loss}
L_{\text{square}}(m) = \left(m - 1\right)^2
\end{equation}

However, there are two drawbacks of square loss when applied to binary classification. First, in binary classification it does not make sense to penalize large positive margins. Second, as mentioned earlier, square loss has the same flaw as all convex losses---penalties for large negative margins grow unboundedly, which can cause non-robustness with respect to label noise. 

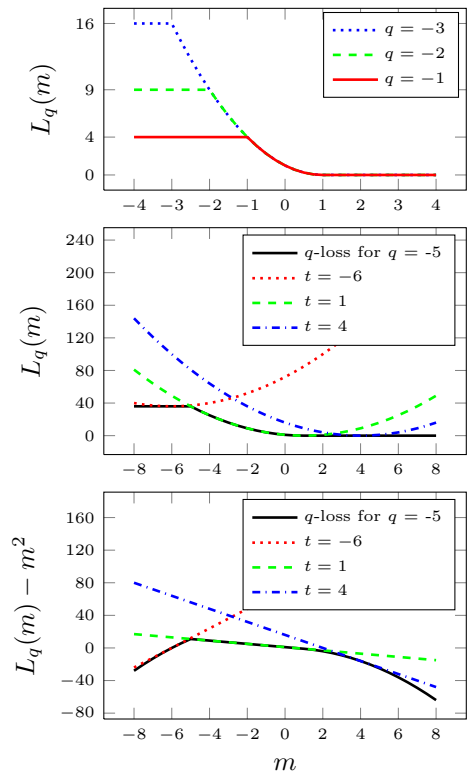
\begin{figure}[t]
\centering
\begin{tikzpicture}
	\begin{axis}[legend style={font=\tiny},name=plot1,tick label style={font=\tiny},xtick={-4,-3,-2,-1,0,1,2,3,4},ytick={0,4,9,16},ylabel=$L_q(m)$,height=4cm,width=6.4cm]
	\addplot[mark=none,color=blue,style=dotted,line width=1pt] coordinates {(-4,16) (-3.75,16) (-3.5,16) (-3.25,16) (-3,16) (-2.75,14.0625) (-2.5,12.25) (-2.25,10.5625) (-2,9) (-1.75,7.5625) (-1.5,6.25) (-1.25,5.0625) (-1,4) (-0.75,3.0625) (-0.5,2.25) (-0.25,1.5625) (0,1) (0.25,0.5625) (0.5,0.25) (0.75,0.0625) (1,0) (1.25,0) (1.5,0) (1.75,0) (2,0) (2.25,0) (2.5,0) (2.75,0) (3,0) (3.25,0) (3.5,0) (3.75,0) (4,0)};
	\addlegendentry{$q = -3$}
	\addplot[mark=none,color=green,style=dashed,,line width=1pt] coordinates {(-4,9) (-3.75,9) (-3.5,9) (-3.25,9) (-3,9) (-2.75,9) (-2.5,9) (-2.25,9) (-2,9) (-1.75,7.5625) (-1.5,6.25) (-1.25,5.0625) (-1,4) (-0.75,3.0625) (-0.5,2.25) (-0.25,1.5625) (0,1) (0.25,0.5625) (0.5,0.25) (0.75,0.0625) (1,0) (1.25,0) (1.5,0) (1.75,0) (2,0) (2.25,0) (2.5,0) (2.75,0) (3,0) (3.25,0) (3.5,0) (3.75,0) (4,0)};
	\addlegendentry{$q = -2$}
	\addplot[mark=none,color=red,,line width=1pt] coordinates {(-4,4) (-3.75,4) (-3.5,4) (-3.25,4) (-3,4) (-2.75,4) (-2.5,4) (-2.25,4) (-2,4) (-1.75,4) (-1.5,4) (-1.25,4) (-1,4) (-0.75,3.0625) (-0.5,2.25) (-0.25,1.5625) (0,1) (0.25,0.5625) (0.5,0.25) (0.75,0.0625) (1,0) (1.25,0) (1.5,0) (1.75,0) (2,0) (2.25,0) (2.5,0) (2.75,0) (3,0) (3.25,0) (3.5,0) (3.75,0) (4,0)};
	\addlegendentry{$q = -1$}
 	\end{axis}
	\begin{axis}[legend style={font=\tiny},name=plot2,at={($(plot1.south)-(0,0.5cm)$)},anchor=north,tick label style={font=\tiny},xtick={-8,-6,-4,-2,0,2,4,6,8},ytick={0,40,80,120,160,200,240},ylabel=$L_q(m)$,height=4.6cm,width=6.4cm,legend style={
cells={anchor=west}, legend pos=north east}]
	\addplot[mark=none,color=black,line width=1pt] coordinates {(-8.000000,36.000000) (-7.500000,36.000000) (-7.000000,36.000000)
(-6.500000,36.000000) (-6.000000,36.000000) (-5.500000,36.000000)
(-5.000000,36.000000) (-4.500000,30.250000) (-4.000000,25.000000)
(-3.500000,20.250000) (-3.000000,16.000000) (-2.500000,12.250000)
(-2.000000,9.000000) (-1.500000,6.250000) (-1.000000,4.000000)
(-0.500000,2.250000) (0.000000,1.000000) (0.500000,0.250000)
(1.000000,0.000000) (1.500000,0.000000) (2.000000,0.000000)
(2.500000,0.000000) (3.000000,0.000000) (3.500000,0.000000)
(4.000000,0.000000) (4.500000,0.000000) (5.000000,0.000000)
(5.500000,0.000000) (6.000000,0.000000) (6.500000,0.000000)
(7.000000,0.000000) (7.500000,0.000000) (8.000000,0.000000)};
	\addlegendentry{$q$-loss for $q$ = -5}
	\addplot[mark=none,color=red,style=dotted,line width=1pt] coordinates {(-8.000000,40.000000) (-7.500000,38.250000) (-7.000000,37.000000)
(-6.500000,36.250000) (-6.000000,36.000000) (-5.500000,36.250000)
(-5.000000,37.000000) (-4.500000,38.250000) (-4.000000,40.000000)
(-3.500000,42.250000) (-3.000000,45.000000) (-2.500000,48.250000)
(-2.000000,52.000000) (-1.500000,56.250000) (-1.000000,61.000000)
(-0.500000,66.250000) (0.000000,72.000000) (0.500000,78.250000)
(1.000000,85.000000) (1.500000,92.250000) (2.000000,100.000000)
(2.500000,108.250000) (3.000000,117.000000) (3.500000,126.250000)
(4.000000,136.000000) (4.500000,146.250000) (5.000000,157.000000)
(5.500000,168.250000) (6.000000,180.000000) (6.500000,192.250000)
(7.000000,205.000000) (7.500000,218.250000) (8.000000,232.000000)};
	\addlegendentry{$t = -6$}
	\addplot[mark=none,color=green,style=dashed,line width=1pt] coordinates {(-8.000000,81.000000) (-7.500000,72.250000) (-7.000000,64.000000)
(-6.500000,56.250000) (-6.000000,49.000000) (-5.500000,42.250000)
(-5.000000,36.000000) (-4.500000,30.250000) (-4.000000,25.000000)
(-3.500000,20.250000) (-3.000000,16.000000) (-2.500000,12.250000)
(-2.000000,9.000000) (-1.500000,6.250000) (-1.000000,4.000000)
(-0.500000,2.250000) (0.000000,1.000000) (0.500000,0.250000)
(1.000000,0.000000) (1.500000,0.250000) (2.000000,1.000000)
(2.500000,2.250000) (3.000000,4.000000) (3.500000,6.250000)
(4.000000,9.000000) (4.500000,12.250000) (5.000000,16.000000)
(5.500000,20.250000) (6.000000,25.000000) (6.500000,30.250000)
(7.000000,36.000000) (7.500000,42.250000) (8.000000,49.000000)};
	\addlegendentry{$t = 1$}
		\addplot[mark=none,color=blue,style=dashdotted,line width=1pt] coordinates {(-8.000000,144.000000) (-7.500000,132.250000) (-7.000000,121.000000)
(-6.500000,110.250000) (-6.000000,100.000000) (-5.500000,90.250000)
(-5.000000,81.000000) (-4.500000,72.250000) (-4.000000,64.000000)
(-3.500000,56.250000) (-3.000000,49.000000) (-2.500000,42.250000)
(-2.000000,36.000000) (-1.500000,30.250000) (-1.000000,25.000000)
(-0.500000,20.250000) (0.000000,16.000000) (0.500000,12.250000)
(1.000000,9.000000) (1.500000,6.250000) (2.000000,4.000000)
(2.500000,2.250000) (3.000000,1.000000) (3.500000,0.250000)
(4.000000,0.000000) (4.500000,0.250000) (5.000000,1.000000)
(5.500000,2.250000) (6.000000,4.000000) (6.500000,6.250000)
(7.000000,9.000000) (7.500000,12.250000) (8.000000,16.000000)};
	\addlegendentry{$t = 4$}
 	\end{axis}
	\begin{axis}[legend style={font=\tiny},name=plot3,at={($(plot2.south)-(0,0.5cm)$)},anchor=north,tick label style={font=\tiny},xtick={-8,-6,-4,-2,0,2,4,6,8},ytick={-80,-40,0,40,80,120,160,200,240},xlabel=$m$,ylabel=$L_q(m) - m^2$,height=4.6cm,width=6.4cm,legend style={
cells={anchor=west}, legend pos=north east}]
	\addplot[mark=none,color=black,line width=1pt] coordinates {(-8.000000,-28.000000) (-7.500000,-20.250000) (-7.000000,-13.000000)
(-6.500000,-6.250000) (-6.000000,0.000000) (-5.500000,5.750000)
(-5.000000,11.000000) (-4.500000,10.000000) (-4.000000,9.000000)
(-3.500000,8.000000) (-3.000000,7.000000) (-2.500000,6.000000)
(-2.000000,5.000000) (-1.500000,4.000000) (-1.000000,3.000000)
(-0.500000,2.000000) (0.000000,1.000000) (0.500000,0.000000)
(1.000000,-1.000000) (1.500000,-2.250000) (2.000000,-4.000000)
(2.500000,-6.250000) (3.000000,-9.000000) (3.500000,-12.250000)
(4.000000,-16.000000) (4.500000,-20.250000) (5.000000,-25.000000)
(5.500000,-30.250000) (6.000000,-36.000000) (6.500000,-42.250000)
(7.000000,-49.000000) (7.500000,-56.250000) (8.000000,-64.000000)};
	\addlegendentry{$q$-loss for $q$ = -5}
	\addplot[mark=none,color=red,style=dotted,line width=1pt] coordinates {(-8.000000,-24.000000) (-7.500000,-18.000000) (-7.000000,-12.000000)
(-6.500000,-6.000000) (-6.000000,0.000000) (-5.500000,6.000000)
(-5.000000,12.000000) (-4.500000,18.000000) (-4.000000,24.000000)
(-3.500000,30.000000) (-3.000000,36.000000) (-2.500000,42.000000)
(-2.000000,48.000000) (-1.500000,54.000000) (-1.000000,60.000000)
(-0.500000,66.000000) (0.000000,72.000000) (0.500000,78.000000)
(1.000000,84.000000) (1.500000,90.000000) (2.000000,96.000000)
(2.500000,102.000000) (3.000000,108.000000) (3.500000,114.000000)
(4.000000,120.000000) (4.500000,126.000000) (5.000000,132.000000)
(5.500000,138.000000) (6.000000,144.000000) (6.500000,150.000000)
(7.000000,156.000000) (7.500000,162.000000) (8.000000,168.000000)};
	\addlegendentry{$t = -6$}
	\addplot[mark=none,color=green,style=dashed,line width=1pt] coordinates {(-8.000000,17.000000) (-7.500000,16.000000) (-7.000000,15.000000)
(-6.500000,14.000000) (-6.000000,13.000000) (-5.500000,12.000000)
(-5.000000,11.000000) (-4.500000,10.000000) (-4.000000,9.000000)
(-3.500000,8.000000) (-3.000000,7.000000) (-2.500000,6.000000)
(-2.000000,5.000000) (-1.500000,4.000000) (-1.000000,3.000000)
(-0.500000,2.000000) (0.000000,1.000000) (0.500000,0.000000)
(1.000000,-1.000000) (1.500000,-2.000000) (2.000000,-3.000000)
(2.500000,-4.000000) (3.000000,-5.000000) (3.500000,-6.000000)
(4.000000,-7.000000) (4.500000,-8.000000) (5.000000,-9.000000)
(5.500000,-10.000000) (6.000000,-11.000000) (6.500000,-12.000000)
(7.000000,-13.000000) (7.500000,-14.000000) (8.000000,-15.000000)};
	\addlegendentry{$t = 1$}
		\addplot[mark=none,color=blue,style=dashdotted,line width=1pt] coordinates {(-8.000000,80.000000) (-7.500000,76.000000) (-7.000000,72.000000)
(-6.500000,68.000000) (-6.000000,64.000000) (-5.500000,60.000000)
(-5.000000,56.000000) (-4.500000,52.000000) (-4.000000,48.000000)
(-3.500000,44.000000) (-3.000000,40.000000) (-2.500000,36.000000)
(-2.000000,32.000000) (-1.500000,28.000000) (-1.000000,24.000000)
(-0.500000,20.000000) (0.000000,16.000000) (0.500000,12.000000)
(1.000000,8.000000) (1.500000,4.000000) (2.000000,0.000000)
(2.500000,-4.000000) (3.000000,-8.000000) (3.500000,-12.000000)
(4.000000,-16.000000) (4.500000,-20.000000) (5.000000,-24.000000)
(5.500000,-28.000000) (6.000000,-32.000000) (6.500000,-36.000000)
(7.000000,-40.000000) (7.500000,-44.000000) (8.000000,-48.000000)};
	\addlegendentry{$t = 4$}
 	\end{axis}
\end{tikzpicture}
\caption{\emph{Top}: $q$-loss for different values of $q$. \emph{Middle}: $q$-loss with three members of the quadratic upper bounds family. $t \in \mathbb{R}$ is the variational parameter. \emph{Bottom}: Tranforming the $y$-axis for concavity.}
\label{fig:$q$-loss}
\end{figure}
With these considerations in mind, we modify square loss in order to obtain a training formulation for binary classification that is both compatible with quantum hardware and robust to label noise. The resulting loss, which we name \emph{q-loss} (Fig.~\ref{fig:$q$-loss}, top), is essentially a doubly truncated version of \eqref{square_loss} with parameterization over $q \in (-\infty, 0]$ defined as follows: 
\begin{definition}[q-loss]
\begin{equation}\label{q_loss_direct}
L_{q}(m) = \min\left(\left(1 - q\right)^2, \left(\max\left(0, 1 - m\right)\right)^2\right)
\end{equation}
\end{definition}

Unfortunately, \eqref{q_loss_direct} does not lead to a QUBO. However, it turns out that we can transform it into a problem which can be solved as a QUBO. The basic idea is to find a variational approximation via a family of quadratic functions that upper-bound $q$-loss and are governed by a variational parameter $t \in \mathbb{R}$ as shown in Fig.~\ref{fig:$q$-loss}, middle.

\begin{theorem}\label{q_loss_theorem}
q-loss in \eqref{q_loss_direct} is equivalent to: 
\begin{align}
&L_q\left(m\right) = \min_t\left\{\left(m - t\right)^2 + \left(1 - q\right)^2\frac{\left(1 - \sign\left(t - 1\right)\right)}{2}\right\}\nonumber
\end{align}
\end{theorem}

\begin{proof}
Since $q$-loss is non-convex, the standard derivation via convex duality \cite{Jaakkola2} dictates that we first find a new coordinate system in which $q$-loss is concave or convex. Then we calculate the conjugate function for linear bounds in the transformed space and transform back to the original space where the linear bounds become the quadratic bounds shown in Fig.~\ref{fig:$q$-loss}, middle. Because of the presence of two constant segments in $q$-loss, any coordinate system in which the two axes are independent transformations of the original $x$ and $y$ axes clearly cannot result in concavity or convexity. Thereby we are led to the transformation $f(y) = y - x^2$, which gives $f\left(L_q\left(m\right)\right) = L_q\left(m\right) - m^2$. It can be seen (Fig.~\ref{fig:$q$-loss}, bottom) that in this transformed space $q$-loss is concave and the quadratic upper bounds become tangent lines. The conjugate function in the transformed space is $g\left(\eta\right) = \min_m\left\{ \eta m - f\left(L_q\left(m\right)\right)\right\}$.

To minimize, we seek stationary points by differentiating $\phi(\eta, m) = \eta m - f(L_q(m))$ with respect to $m$:
\begin{align}
\frac{\partial}{\partial m}\phi\left(\eta, m\right) &= \eta - \frac{d}{d m}L_q\left(m\right) + 2m\\
&=
\begin{cases}
 \eta + 2 & \text{for }m \in (q, 1)\\
 \eta + 2m & \text{for }m \in (-\infty, q) \cup (1, \infty) \hbox{ ,}
\end{cases}\nonumber
\end{align}
as yielded by piecewise differentiation of $L_q(m)$. Setting to $0$ gives the stationary points
\begin{align}
&\eta = - 2 \text{ for }m \in (q, 1)\\
&m = -\eta/2 \text{ for }m \in (-\infty, q) \cup (1, \infty) \hbox{ .}
\end{align}
Plugging them back into the conjugate function yields
\begin{align}
g\left(\eta\right) &= 
\begin{cases} \nonumber
 -\frac{\eta^2}{4} - (1 - q)^2& \text{for }m \in (-\infty, q)\\
 -1 & \text{for }m \in (q, 1)\\
-\frac{\eta^2}{4}& \text{for }m \in (1, \infty)\\
\end{cases}\\
&= -\frac{\eta^2}{4} - (1 - q)^2\frac{\left(1 - \sign\left(-\frac{\eta}{2} - 1\right)\right)}{2} \hbox{ .}
\end{align}
In accordance with convex duality,
\begin{align}
&f\left(L_q\left(m\right)\right) = \min_{\eta} \left\{\eta m - g\left(\eta\right)\right\}\\
&= \min_{\eta} \left\{\eta m + \frac{\eta^2}{4} + (1 - q)^2\frac{\left(1 - \sign\left(-\frac{\eta}{2} - 1\right)\right)}{2}\right\}\text{ .}\nonumber
\end{align}
Transforming back into the original space and setting $t = -\eta/2$, the variational upper bound for $q$-loss is
\begin{align}\label{variational_upper_bound}
&L_q\left(m\right) = f^{-1}\left(f\left(L_q\left(m\right)\right)\right)\\
& = \min_t\left\{\left(m - t\right)^2 + \left(1 - q\right)^2\frac{\left(1 - \sign\left(t - 1\right)\right)}{2}\right\}\nonumber
\end{align}
\end{proof}

\subsection{Latent variables view}\label{latent_variables_view}
Traditionally when facing non-convex optimization problems, a viable approach is to introduce latent variables that allow reformulating over a simpler family of functions. This is precisely what Theorem~\ref{q_loss_theorem} achieves. For any fixed $m$, the latent variable $t \in \mathbb{R}$ gives a convex optimization problem whose minimum is $L_{q}(m)$: 
\begin{align}\label{q_loss}
&L_{q}\left(m\right) = h\left(m, t^*\left(m\right)\right)\text{ , where}\\
&t^*\left(m\right) = \arg \min_{t}\left\{h\left(m, t\right)\right\}\nonumber \\
&h\left(m, t\right) = \left(m - t\right)^2  + \left(1 - q\right)^2 \left(1 - \sign\left(t - 1 \right)\right)/2\nonumber
\end{align}

The regularized risk minimization \eqref{train_opt} with empirical risk over $L_{q}$ in the form \eqref{q_loss} is amenable to a block coordinate descent method for jointly optimizing the model parameters $(\vc{w}, b)$ and the latent variables $t_s$ for $s = 1, \ldots, S$: similarly to EM, alternate between convex optimization runs over the latent variables (\emph{$t$ step}) and the model parameters (\emph{$w$ step}). Even though such methods do well on some problems with certain benign structure---e.g. Gaussian mixtures \cite{Dempster})---they are also known to fail on other problems that lack such structure. We believe $q$-loss belongs to the latter group and have verified that a block coordinate descent method is likely to be sensitive to initialization and is quickly terminating in bad local minima. The intuitive reason is that due to the quadratically growing penalty for mismatching a margin with its latent variable, the $t$ step tends to lock in the model parameters found during the previous $w$ step, thus possibly preventing the next $w$ step from moving to a different model. The impact of this effect becomes ever more severe for large data with $S >> N$.

On the other hand, by transforming \eqref{q_loss_direct} into \eqref{q_loss} we have made training with $q$-loss representable in QUBO form albeit at the expense of additional variables. Section~\ref{explicit_qubo} of the supplementary material explicitly shows the QUBO problem that can be derived from \eqref{q_loss}. Since the goal of AQO is to perform global optimization simultaneously over all variables, we believe AQO is a much better candidate for training with $q$-loss. Besides making the QUBO formulation possible, the introduction of latent variables also gives rise to an intuitive interpretation of the mechanism by which $q$-loss achieves robustness when compared to the non-robustness of square loss. While in \eqref{square_loss} the fixed target $1$ has to be matched as closely as possible by $m$, in \eqref{q_loss} $t$ plays the role of a flexible target that can change sign for a large negative margin, thereby flagging that training example as mislabeled. For any $m$, the minimizer $t^*(m)$ in \eqref{q_loss} belongs to one of three cases:
\vspace{0.1cm}\\
\begin{tabular}{ c c l c c l }
	& $\bullet$ & Case \Rmnum{1}: & $m \geq 1$ & $\Rightarrow$ & $t^*(m) = m$ \\
	& $\bullet$ & Case \Rmnum{2}: & $q < m < 1$ & $\Rightarrow$ & $t^*(m) = 1$ \\
	& $\bullet$ & Case  \Rmnum{3}: & $m \leq q$ & $\Rightarrow$ & $t^*(m) = m$\\
\end{tabular}
\vspace{0.1cm}\\
Case \Rmnum{1} ensures zero penalty for large positive margins; Case \Rmnum{2} produces the same quadratic penalty as \eqref{square_loss}; Case \Rmnum{3} can be seen as flipping the label of a possibly mislabeled example but also incurring a constant penalty of $(1 - q)^2$ in order to not lose connection with the original labeling. Thus, the hyper-parameter $q$ defines the largest negative margin to be tolerated. A training example that has a negative margin with some larger magnitude gets flipped with constant penalty.

\section{Bounding $q$}\label{bounding_q}
While it is difficult to formalize any general statements about the computational hardness of $q$-loss, it is easily recognized that the hardness depends on the size of the parabolic segment controlled by $q$. For $q\rightarrow -\infty$, even the negative margins of highest magnitude incur the usual quadratic penalty, and the loss becomes effectively convex. For smaller $q$ the loss becomes similar to 0-1 loss, so the resulting optimization problems may be approaching the hardness of the corresponding 0-1 loss problems. However, the most beneficial regime of operation is not known a-priori. This necessitates cross-validation over $q$, which, depending on the noise level, we expect to result in some trade-off between robustness and computational hardness. For the purpose of choosing values for cross-validation, we give an approximate lower bound for $q$ as a function of our estimate of the underlying Bayes error in the data and the label noise that we might artificially insert into the training set for robustness evaluation.

Let the effective Bayes error be $\beta_{eff} \in [0, 0.5)$. This should account both for the Bayes error $\beta_0$ of the data that we are given and the additional error $\nu \in [0, 0.5)$ that we introduce by injecting label noise. Then if we wish for the entire $\beta_{eff}$ portion of the training set to be flagged by $q$-loss as mislabeled, the empirical risk is $R(\vc{w},b) \geq \beta_{eff}*(1-q)^2$. But we know the trivial solution consisting of all $0$ weights has $R(\vc{0},0) = 1$. Then we want $\beta_{eff}*(1-q)^2 < 1$, which, together with $q \in (-\infty, 0]$, gives $q \in (1 - 1/\sqrt{\beta_{eff}}, 0]$.

Usually we do not have $\beta_0$, but we can obtain an empirical estimate by training on the given data: $\beta_{emp} = \beta_0 + \beta_{opt} + \beta_{gen}$, where $\beta_{opt}$ is the additional error caused by imperfect optimization, and $\beta_{gen}$ represents the generalization component of the overall test error. Assuming $\beta_{emp}$ is sufficiently close to $\beta_0$ and accounting for the artificially introduced label noise $\nu$, we set $\beta_{eff} = \beta_{emp} - 2\beta_{emp}\nu + \nu$. The subtraction corrects for originally bad examples that flip under $\nu$.

\section{Low-precision discrete variables}\label{discrete_weights}
The quantum optimization processor that we aim to deploy for solving $q$-loss requires problems to be discrete and formulated as QUBO. Further, the current hardware can handle a maximum of $512$ binary variables, which imposes the additional requirement of being frugal with the bit-depth of weights. To that end we discretize the elements of $\vc{w}$ to some low bit-depth $d_{w} < 64$. While this approach is somewhat unconventional, \cite{Neven08b} argued there is no fundamental reason why the weights should need high precision and in fact showed a favorable sufficiency condition of $d_{w} \approx \log(S/N)$ in the case of binary features. Even though classifiers constructed out of more general features have not been studied in this way, our experiments provide support for using low-precision weights.

The reason for fixing at $1$ the smallest positive margin that yields zero penalty in $q$-loss is the same as in hinge loss SVM \cite{Bishop}: any arbitrary rescaling of the weights $\vc{w} \rightarrow \kappa\vc{w}$ and bias $b \rightarrow \kappa b$ does not change the geometric distance $y_s\left(\vc{w}^T\vc{x}_s + b\right)/\norm{\vc{w}}$ from a data point $(\vc{x}_s, y_s)$ to the decision surface. Therefore, we can assume a margin of $1$ for the correctly classified point that is closest to the decision surface. However, this freedom of arbitrary rescaling becomes complicated when the bit-depth of weights is lowered. We want the intervals for weight variables to cover the maximum magnitude that the interplay between margin enforcement and regularization may demand. On the other hand, a loose interval decreases the effective precision in sub-intervals that may really matter. For that reason we derive a $\lambda$-dependent bound for setting the intervals in which discrete weight variables can take values.

Let $F(\vc{w}, b) = R(\vc{w}, b) + \lambda \Omega(\vc{w})$ be the objective function. For $q$-loss, $F(\vc{0}, 0) = R(\vc{0}, 0) = 1$ and $\exists~\hat{\vc{w}} \ni F(\vc{0},0) = \lambda \Omega(\hat{\vc{w}})$. Then,
\begin{equation}
F(\hat{\vc{w}},b) = R(\hat{\vc{w}},b) + \lambda \Omega(\hat{\vc{w}}) \geq \lambda \Omega(\hat{\vc{w}}) = F(\vc{0},0)\\
\end{equation}
Also, $F(-\hat{\vc{w}},b) \geq F(\vc{0},0)$ because $\Omega(-\hat{\vc{w}}) = \Omega(\hat{\vc{w}})$. Now consider any $\tilde{\vc{w}} \ni \Omega(\tilde{\vc{w}}) \geq \Omega(\hat{\vc{w}})$:
\begin{equation}
F(\tilde{\vc{w}},b) \geq \lambda \Omega(\tilde{\vc{w}}) \geq \lambda \Omega(\hat{\vc{w}}) = F(\vc{0},0)
\end{equation}
Hence, $F(\tilde{\vc{w}},b) \geq F(\hat{\vc{w}},b) \geq F(\vc{0},0)$ and $F(-\tilde{\vc{w}},b) \geq F(-\hat{\vc{w}},b) \geq F(\vc{0},0)$. Thus, we can use $\Omega(\hat{\vc{w}}) = 1/\lambda$ to bound the intervals in which the weight variables live while ensuring that the minimizer of $F$ belongs to these intervals. For $\ell_{2}$-norm regularization, $\Omega(\hat{\vc{w}}) = \norm{\hat{\vc{w}}}_2 \geq \norm{\hat{\vc{w}}}_\infty = \max(\abs{\hat{\vc{w}}})$, so we train by optimizing $w_j~\forall j$ only in the interval $[-\Omega(\hat{\vc{w}}), \Omega(\hat{\vc{w}})]$.

The discrete optimization problem for training with $q$-loss and $\ell_{2}$-norm regularization is:
\begin{equation}\label{opt_discretized}
(\dot{\vc{w}}, \dot{b})^{*} = \arg \min_{\dot{\vc{w}}, \dot{b}} \left\{\frac{1}{S}\sum_{s = 1}^S L_{q}\left(y_s(\dot{\vc{w}}^T\vc{x}_s + \dot{b})\right) + \lambda \norm{\dot{\vc{w}}}_2\right\} \hbox{,}
\end{equation}
where $\dot{\vc{w}}$ and $\dot{b}$ are the discretized $\vc{w}$ and $b$, and $\lambda \in \mathbb{R}_{\geq0}$ controls the relative importance of regularization.

\section{Experimental evaluation}\label{experiments}
While prior work on non-convex losses applied various forms of convex optimization \cite{Hamed, Yuille, lbfgs} hoping they can still be solved with somewhat reasonable quality, we take the approach of directly tackling the resulting problems by discrete optimization. Admittedly, this choice makes the optimization method largely oblivious to existing benign structure and may cause us to face NP-hardness in certain situations. However, we do this for the purpose of being compatible with emerging quantum hardware that can be employed as a black-box discrete optimization engine having the potential to do well on such problems. 

Quantum hardware was already successfully deployed by \cite{CarDetector} on a large-scale training problem with square loss and $\ell_0$-norm regularization. In the present work on $q$-loss with $\ell_2$-norm regularization, we only verify the validity of our approach by using Tabu search \cite{Palubeckis2004} as a classical heuristic stand-in and leave the quantum hardware to future work. A quantum optimization with $q$-loss is expected to achieve in shorter time equal or better results than our classical optimization setup. We do not report CPU time comparisons because they are irrelevant in the absence of quantum hardware runs.
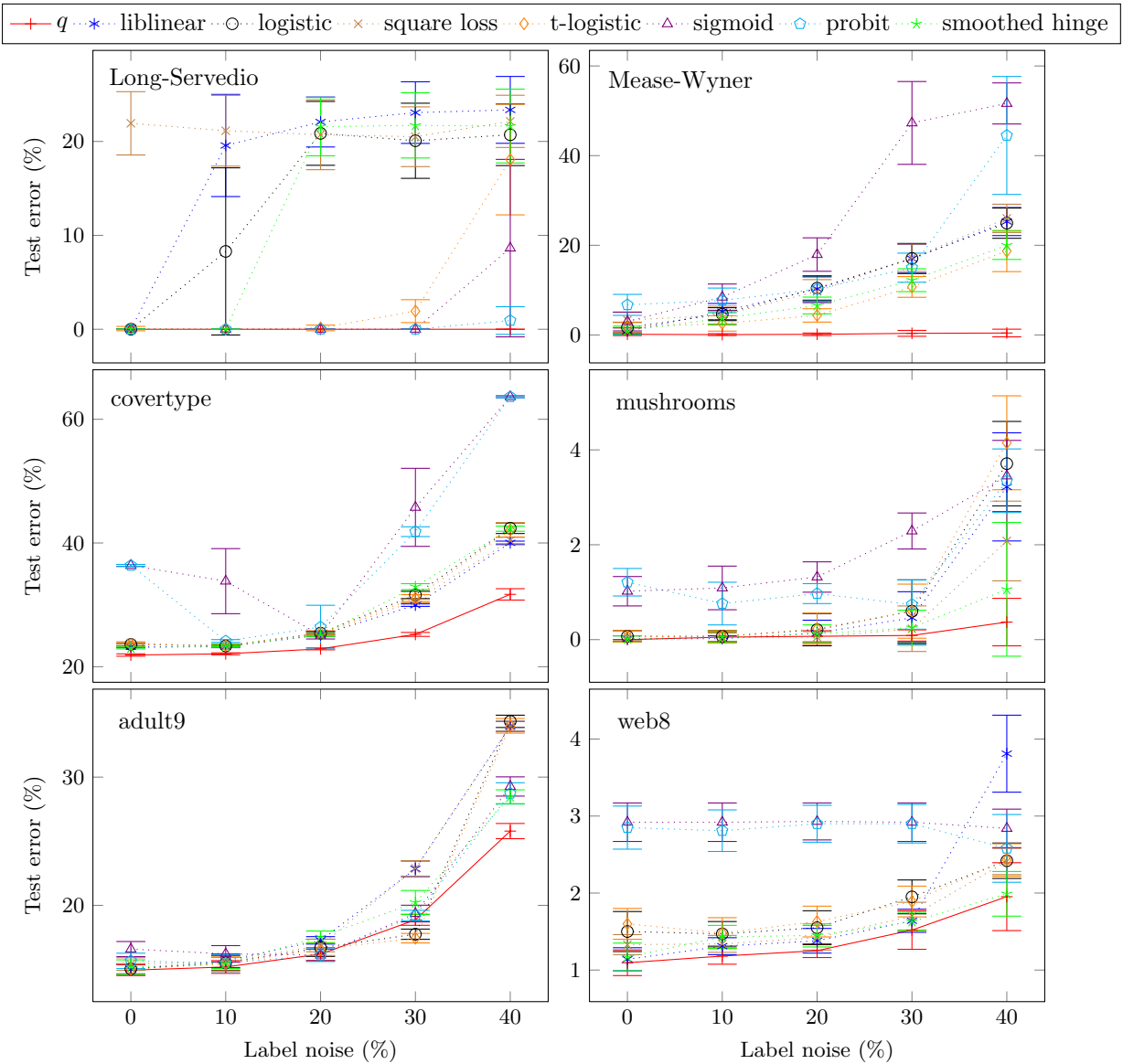
\begin{figure*}[t!]
\centering
\tikzset{every mark/.append style={solid,scale=1.2}}
\pgfplotsset{every axis/.append style={line width=0.4pt,xtick = {0,10,20,30,40},xticklabels={},height=6.13cm,width=8.2cm,cycle list ={
{red,solid,mark=+},
{blue,dotted,mark=asterisk},
{black,dotted,mark=o},
{brown,dotted,mark=x},
{orange,dotted,mark=diamond},
{violet,dotted,mark=triangle},
{cyan,dotted,mark=pentagon},
{green,dotted,mark=star}}},
every error bar/.append style={solid}}
\begin{tikzpicture}                             
	\begin{axis} [name=plot1,legend columns=-1,legend entries={$q$,liblinear,logistic,square loss,t-logistic\newline,sigmoid,probit,smoothed hinge},
	legend style={at={(-0.2,1.02)}, anchor=south west},tick label style={font=\small},y label style={font=\small},ylabel=Test error ($\%$)]
	\node at (axis description cs:0.2,0.91) {Long-Servedio};
	\addplot+[error bars/.cd, y dir=both, y explicit, error mark=|,error mark options={scale=3.0}]
		coordinates {
(0,0.000000)+-(0,0.000000)
(10,0.000000)+-(0,0.000000)
(20,0.000000)+-(0,0.000000)
(30,0.000000)+-(0,0.000000)
(40,0.000000)+-(0,0.000000)
		};

	\addplot+[error bars/.cd, y dir=both, y explicit, error mark=|,error mark options={scale=3.0}]
		coordinates {
			(0,0) +- (0,0)
			(10,19.57) +- (0,5.44)
			(20,22.07) +- (0,2.66)
			(30,23.07) +- (0,3.28)
			(40,23.36) +- (0,3.56)
		};

	\addplot+[error bars/.cd, y dir=both, y explicit, error mark=|,error mark options={scale=3.0}]
		coordinates {
			(0,0) +- (0,0)
			(10,8.29) +- (0,8.9)
			(20,20.86) +- (0,3.4)
			(30,20.07) +- (0,4)
			(40,20.71) +- (0,3.28)
		};

	\addplot+[error bars/.cd, y dir=both, y explicit, error mark=|,error mark options={scale=3.0}]
		coordinates {
			(0,21.93) +- (0,3.37)
			(10,21.14) +- (0,3.77)
			(20,20.71) +- (0,3.72)
			(30,20.5) +- (0,3.18)
			(40,22.14) +- (0,2.78)
		};

	\addplot+[error bars/.cd, y dir=both, y explicit, error mark=|,error mark options={scale=3.0}]
		coordinates {
			(0,0.07) +- (0,0.23)
			(10,0) +- (0,0)
			(20,0.14) +- (0,0.3)
			(30,1.93) +- (0,1.22)
			(40,18.07) +- (0,5.89)
		};

	\addplot+[error bars/.cd, y dir=both, y explicit, error mark=|,error mark options={scale=3.0}]
		coordinates {
			(0,0) +- (0,0)
			(10,0) +- (0,0)
			(20,0) +- (0,0)
			(30,0) +- (0,0)
			(40,8.64) +- (0,9.45)
		};

	\addplot+[error bars/.cd, y dir=both, y explicit, error mark=|,error mark options={scale=3.0}]
		coordinates {
			(0,0) +- (0,0)
			(10,0) +- (0,0)
			(20,0) +- (0,0)
			(30,0) +- (0,0)
			(40,0.93) +- (0,1.47)
		};

	\addplot+[error bars/.cd, y dir=both, y explicit, error mark=|,error mark options={scale=3.0}]
		coordinates {
			(0,0) +- (0,0)
			(10,0) +- (0,0)
			(20,21.57) +- (0,3.1)
			(30,21.71) +- (0,3.47)
			(40,21.64) +- (0,3.93)
		};
	\end{axis}
	\begin{axis} [name=plot2,at={($(plot1.east)+(0.6cm,0)$)},anchor=west,tick label style={font=\small}]
	\node at (axis description cs:0.2,0.9) {Mease-Wyner};
	\addplot+[error bars/.cd, y dir=both, y explicit, error mark=|,error mark options={scale=3.0}]
		coordinates {
(0,0.142900)+-(0,0.285700)
(10,0.071400)+-(0,0.214300)
(20,0.142900)+-(0,0.285700)
(30,0.357100)+-(0,0.658500)
(40,0.428600)+-(0,0.857100)
		};

	\addplot+[error bars/.cd, y dir=both, y explicit, error mark=|,error mark options={scale=3.0}]
		coordinates {
			(0,0.86) +- (0,0.74)
			(10,5.21) +- (0,1.81)
			(20,10.29) +- (0,2.9)
			(30,17) +- (0,3.24)
			(40,25.29) +- (0,3.13)
		};

	\addplot+[error bars/.cd, y dir=both, y explicit, error mark=|,error mark options={scale=3.0}]
		coordinates {
			(0,1.64) +- (0,1.17)
			(10,4.64) +- (0,1.44)
			(20,10.43) +- (0,2.66)
			(30,17.07) +- (0,3.34)
			(40,24.93) +- (0,3.37)
		};

	\addplot+[error bars/.cd, y dir=both, y explicit, error mark=|,error mark options={scale=3.0}]
		coordinates {
			(0,3.29) +- (0,1.76)
			(10,4.43) +- (0,1.99)
			(20,9.71) +- (0,2.63)
			(30,17.21) +- (0,3.04)
			(40,26) +- (0,3.16)
		};

	\addplot+[error bars/.cd, y dir=both, y explicit, error mark=|,error mark options={scale=3.0}]
		coordinates {
			(0,1.64) +- (0,1.22)
			(10,2.57) +- (0,1.72)
			(20,4.36) +- (0,1.49)
			(30,10.71) +- (0,2.31)
			(40,18.71) +- (0,4.58)
		};

	\addplot+[error bars/.cd, y dir=both, y explicit, error mark=|,error mark options={scale=3.0}]
		coordinates {
			(0,3) +- (0,2.1)
			(10,8.43) +- (0,2.93)
			(20,17.93) +- (0,3.73)
			(30,47.29) +- (0,9.24)
			(40,51.64) +- (0,4.58)
		};

	\addplot+[error bars/.cd, y dir=both, y explicit, error mark=|,error mark options={scale=3.0}]
		coordinates {
			(0,6.71) +- (0,2.34)
			(10,7.71) +- (0,2.75)
			(20,10.21) +- (0,2.67)
			(30,15) +- (0,3.25)
			(40,44.5) +- (0,13.16)
		};

	\addplot+[error bars/.cd, y dir=both, y explicit, error mark=|,error mark options={scale=3.0}]
		coordinates {
			(0,1) +- (0,0.96)
			(10,3.71) +- (0,1.5)
			(20,6.57) +- (0,1.87)
			(30,12.21) +- (0,2.55)
			(40,20) +- (0,3.16)
		};
	\end{axis}
	\begin{axis} [name=plot3,at={($(plot1.south)-(0,0.1cm)$)},anchor=north,tick label style={font=\small},y label style={font=\small},ylabel=Test error ($\%$)]
	\node at (axis description cs:0.15,0.9) {covertype};
	\addplot+[error bars/.cd, y dir=both, y explicit, error mark=|,error mark options={scale=3.0}]
		coordinates {
(0,21.886400)+-(0,0.193000)
(10,22.082400)+-(0,0.158800)
(20,22.880500)+-(0,0.153500)
(30,25.220600)+-(0,0.319600)
(40,31.693400)+-(0,0.917000)
		};

	\addplot+[error bars/.cd, y dir=both, y explicit, error mark=|,error mark options={scale=3.0}]
		coordinates {
			(0,23.21) +- (0,0.15)
			(10,23.22) +- (0,0.15)
			(20,25.12) +- (0,0.24)
			(30,29.95) +- (0,0.2)
			(40,40.02) +- (0,0.28)
		};

	\addplot+[error bars/.cd, y dir=both, y explicit, error mark=|,error mark options={scale=3.0}]
		coordinates {
			(0,23.6) +- (0,0.21)
			(10,23.36) +- (0,0.19)
			(20,25.43) +- (0,0.31)
			(30,31.6) +- (0,0.59)
			(40,42.38) +- (0,0.85)
		};

	\addplot+[error bars/.cd, y dir=both, y explicit, error mark=|,error mark options={scale=3.0}]
		coordinates {
			(0,23.85) +- (0,0.17)
			(10,23.27) +- (0,0.14)
			(20,25.27) +- (0,0.36)
			(30,30.63) +- (0,0.28)
			(40,40.4) +- (0,0.51)
		};

	\addplot+[error bars/.cd, y dir=both, y explicit, error mark=|,error mark options={scale=3.0}]
		coordinates {
			(0,23.59) +- (0,0.25)
			(10,23.43) +- (0,0.26)
			(20,25.48) +- (0,0.33)
			(30,31.1) +- (0,0.55)
			(40,42.11) +- (0,1.14)
		};

	\addplot+[error bars/.cd, y dir=both, y explicit, error mark=|,error mark options={scale=3.0}]
		coordinates {
			(0,36.37) +- (0,0.16)
			(10,33.83) +- (0,5.26)
			(20,25.08) +- (0,0.6)
			(30,45.76) +- (0,6.31)
			(40,63.63) +- (0,0.16)
		};

	\addplot+[error bars/.cd, y dir=both, y explicit, error mark=|,error mark options={scale=3.0}]
		coordinates {
			(0,36.37) +- (0,0.16)
			(10,24.15) +- (0,0.23)
			(20,26.4) +- (0,3.53)
			(30,41.83) +- (0,0.79)
			(40,63.63) +- (0,0.16)
		};

	\addplot+[error bars/.cd, y dir=both, y explicit, error mark=|,error mark options={scale=3.0}]
		coordinates {
			(0,23.02) +- (0,0.15)
			(10,23.36) +- (0,0.12)
			(20,25.12) +- (0,0.29)
			(30,32.94) +- (0,0.48)
			(40,42.31) +- (0,0.41)
		};
	\end{axis}
	\begin{axis} [name=plot4,at={($(plot2.south)-(0,0.1cm)$)},anchor=north,tick label style={font=\small}]
	\node at (axis description cs:0.19,0.9) {mushrooms};
	\addplot+[error bars/.cd, y dir=both, y explicit, error mark=|,error mark options={scale=3.0}]
		coordinates {
(0,0.000000)+-(0,0.000000)
(10,0.052800)+-(0,0.112700)
(20,0.070400)+-(0,0.116800)
(30,0.088000)+-(0,0.118100)
(40,0.369700)+-(0,0.500800)
		};

	\addplot+[error bars/.cd, y dir=both, y explicit, error mark=|,error mark options={scale=3.0}]
		coordinates {
			(0,0.02) +- (0,0.06)
			(10,0.02) +- (0,0.06)
			(20,0.14) +- (0,0.27)
			(30,0.46) +- (0,0.55)
			(40,3.22) +- (0,1.14)
		};

	\addplot+[error bars/.cd, y dir=both, y explicit, error mark=|,error mark options={scale=3.0}]
		coordinates {
			(0,0.07) +- (0,0.12)
			(10,0.07) +- (0,0.12)
			(20,0.21) +- (0,0.34)
			(30,0.6) +- (0,0.66)
			(40,3.71) +- (0,0.89)
		};

	\addplot+[error bars/.cd, y dir=both, y explicit, error mark=|,error mark options={scale=3.0}]
		coordinates {
			(0,0.07) +- (0,0.12)
			(10,0.05) +- (0,0.09)
			(20,0.05) +- (0,0.12)
			(30,0.23) +- (0,0.48)
			(40,2.08) +- (0,0.84)
		};

	\addplot+[error bars/.cd, y dir=both, y explicit, error mark=|,error mark options={scale=3.0}]
		coordinates {
			(0,0.07) +- (0,0.12)
			(10,0.07) +- (0,0.12)
			(20,0.23) +- (0,0.33)
			(30,0.6) +- (0,0.57)
			(40,4.15) +- (0,0.99)
		};

	\addplot+[error bars/.cd, y dir=both, y explicit, error mark=|,error mark options={scale=3.0}]
		coordinates {
			(0,1.02) +- (0,0.31)
			(10,1.09) +- (0,0.46)
			(20,1.32) +- (0,0.32)
			(30,2.29) +- (0,0.38)
			(40,3.45) +- (0,0.75)
		};

	\addplot+[error bars/.cd, y dir=both, y explicit, error mark=|,error mark options={scale=3.0}]
		coordinates {
			(0,1.21) +- (0,0.29)
			(10,0.76) +- (0,0.45)
			(20,0.97) +- (0,0.21)
			(30,0.74) +- (0,0.52)
			(40,3.35) +- (0,0.67)
		};

	\addplot+[error bars/.cd, y dir=both, y explicit, error mark=|,error mark options={scale=3.0}]
		coordinates {
			(0,0.02) +- (0,0.06)
			(10,0.05) +- (0,0.12)
			(20,0.12) +- (0,0.19)
			(30,0.25) +- (0,0.37)
			(40,1.06) +- (0,1.41)
		};
	\end{axis}
	\begin{axis} [name=plot5,at={($(plot3.south)-(0,0.1cm)$)},anchor=north,x label style={font=\small},xlabel=Label noise ($\%$),xticklabels={0,10,20,30,40},y label style={font=\small},ylabel=Test error ($\%$),tick label style={font=\small}]
	\node at (axis description cs:0.13,0.9) {adult9};
	\addplot+[error bars/.cd, y dir=both, y explicit, error mark=|,error mark options={scale=3.0}]
		coordinates {
(0,14.945900)+-(0,0.423200)
(10,15.194500)+-(0,0.498600)
(20,16.200600)+-(0,0.496800)
(30,18.871000)+-(0,0.428400)
(40,25.791200)+-(0,0.593100)
		};

	\addplot+[error bars/.cd, y dir=both, y explicit, error mark=|,error mark options={scale=3.0}]
		coordinates {
			(0,15.02) +- (0,0.4)
			(10,15.68) +- (0,0.53)
			(20,17.27) +- (0,0.29)
			(30,22.85) +- (0,0.61)
			(40,33.96) +- (0,0.39)
		};

	\addplot+[error bars/.cd, y dir=both, y explicit, error mark=|,error mark options={scale=3.0}]
		coordinates {
			(0,15.05) +- (0,0.39)
			(10,15.51) +- (0,0.57)
			(20,16.69) +- (0,0.67)
			(30,17.75) +- (0,0.4)
			(40,34.35) +- (0,0.48)
		};

	\addplot+[error bars/.cd, y dir=both, y explicit, error mark=|,error mark options={scale=3.0}]
		coordinates {
			(0,15.42) +- (0,0.49)
			(10,15.6) +- (0,0.51)
			(20,16.87) +- (0,0.49)
			(30,22.88) +- (0,0.6)
			(40,33.97) +- (0,0.39)
		};

	\addplot+[error bars/.cd, y dir=both, y explicit, error mark=|,error mark options={scale=3.0}]
		coordinates {
			(0,15.03) +- (0,0.41)
			(10,15.44) +- (0,0.54)
			(20,16.59) +- (0,0.5)
			(30,17.44) +- (0,0.37)
			(40,34) +- (0,0.57)
		};

	\addplot+[error bars/.cd, y dir=both, y explicit, error mark=|,error mark options={scale=3.0}]
		coordinates {
			(0,16.59) +- (0,0.59)
			(10,16.22) +- (0,0.64)
			(20,16.12) +- (0,0.45)
			(30,19.36) +- (0,0.65)
			(40,29.27) +- (0,0.75)
		};

	\addplot+[error bars/.cd, y dir=both, y explicit, error mark=|,error mark options={scale=3.0}]
		coordinates {
			(0,15.67) +- (0,0.6)
			(10,15.5) +- (0,0.67)
			(20,16.12) +- (0,0.5)
			(30,19.2) +- (0,0.41)
			(40,28.73) +- (0,0.83)
		};

	\addplot+[error bars/.cd, y dir=both, y explicit, error mark=|,error mark options={scale=3.0}]
		coordinates {
			(0,15.19) +- (0,0.56)
			(10,15.5) +- (0,0.43)
			(20,17.53) +- (0,0.47)
			(30,20.22) +- (0,0.94)
			(40,28.47) +- (0,0.54)
		};
	\end{axis}
	\begin{axis} [name=plot6,at={($(plot4.south)-(0,0.1cm)$)},anchor=north,x label style={font=\small},xlabel=Label noise ($\%$),xticklabels={0,10,20,30,40},tick label style={font=\small}]
	\node at (axis description cs:0.12,0.9) {web8};
	\addplot+[error bars/.cd, y dir=both, y explicit, error mark=|,error mark options={scale=3.0}]
		coordinates {
(0,1.094800)+-(0,0.166800)
(10,1.181600)+-(0,0.105700)
(20,1.251500)+-(0,0.089200)
(30,1.519200)+-(0,0.249900)
(40,1.953200)+-(0,0.440700)
		};

	\addplot+[error bars/.cd, y dir=both, y explicit, error mark=|,error mark options={scale=3.0}]
		coordinates {
			(0,1.14) +- (0,0.15)
			(10,1.31) +- (0,0.11)
			(20,1.38) +- (0,0.16)
			(30,1.64) +- (0,0.15)
			(40,3.81) +- (0,0.5)
		};

	\addplot+[error bars/.cd, y dir=both, y explicit, error mark=|,error mark options={scale=3.0}]
		coordinates {
			(0,1.5) +- (0,0.26)
			(10,1.47) +- (0,0.16)
			(20,1.55) +- (0,0.22)
			(30,1.95) +- (0,0.22)
			(40,2.42) +- (0,0.23)
		};

	\addplot+[error bars/.cd, y dir=both, y explicit, error mark=|,error mark options={scale=3.0}]
		coordinates {
			(0,1.33) +- (0,0.13)
			(10,1.34) +- (0,0.11)
			(20,1.42) +- (0,0.16)
			(30,1.7) +- (0,0.18)
			(40,2.41) +- (0,0.17)
		};

	\addplot+[error bars/.cd, y dir=both, y explicit, error mark=|,error mark options={scale=3.0}]
		coordinates {
			(0,1.6) +- (0,0.2)
			(10,1.48) +- (0,0.2)
			(20,1.63) +- (0,0.2)
			(30,1.89) +- (0,0.2)
			(40,2.43) +- (0,0.21)
		};

	\addplot+[error bars/.cd, y dir=both, y explicit, error mark=|,error mark options={scale=3.0}]
		coordinates {
			(0,2.92) +- (0,0.25)
			(10,2.92) +- (0,0.25)
			(20,2.93) +- (0,0.24)
			(30,2.92) +- (0,0.25)
			(40,2.84) +- (0,0.25)
		};

	\addplot+[error bars/.cd, y dir=both, y explicit, error mark=|,error mark options={scale=3.0}]
		coordinates {
			(0,2.85) +- (0,0.28)
			(10,2.81) +- (0,0.27)
			(20,2.9) +- (0,0.24)
			(30,2.9) +- (0,0.25)
			(40,2.58) +- (0,0.44)
		};

	\addplot+[error bars/.cd, y dir=both, y explicit, error mark=|,error mark options={scale=3.0}]
		coordinates {
			(0,1.17) +- (0,0.18)
			(10,1.43) +- (0,0.15)
			(20,1.45) +- (0,0.15)
			(30,1.63) +- (0,0.12)
			(40,1.99) +- (0,0.29)
		};
	\end{axis}
\end{tikzpicture}
 \caption{Test error vs label noise for $8$ methods (see legend) on $2$ synthetic data sets (Long-Servedio and Mease-Wyner) and $4$ UCI  data sets (covertype, mushrooms, adult9, web8). Error bars are obtained from $10$-fold cross-validation. }
\label{fig:error_vs_noise}
\end{figure*}

In order to show robustness, we randomly flip training labels and observe the worsening of test error as a function of increasing label noise. While prior work on robust classification \cite{DingVishy,Bottou2006,Freund09} considered uniform label noise, we note this does not adequately capture the essence of the true mechanism by which label noise trickles into real-world training tasks. In fact, recent work \cite{ManwaniSastry} shows that even convex losses can be robust under uniform noise. Moreover, experience with practical applications confirms that the type of label noise that affects classification accuracy is never independent of the underlying data distribution. For example, if the human taggers preparing training data for a computer vision application receive somewhat inaccurate or ambiguous instructions affecting only one of the classes, the resulting label noise is strongly correlated with that class. For this reason we move to a noise model in which we introduce uniformly random flips only in the labels of one class---here WLOG of the negative class---and keep the labels of the other class clean. In the experiments described below, the percentage label noise refers to the probability with which we flip labels in the negative portion of training data.

We conduct experiments on two synthetic and four UCI data sets (data summary in Section~\ref{data_summary} of supplementary material). The synthetic data sets \cite{LongServedio,MeaWyn07} are designed to provide a stark distinction between robust and non-robust losses. We compare the classification performance of $q$-loss to seven other convex and non-convex $\ell_2$-regularized methods: liblinear ($\ell_2$-loss primal SVM) \cite{liblinear}, t-logistic regression \cite{DingVishy}, smoothed hinge loss \cite{smoothed_hinge}, logistic regression, square loss, sigmoid loss, and probit loss \cite{Bishop}. For all methods except $q$-loss and liblinear we use Petsc/Tao implementations with convex optimization \cite{petsc-web-page,tao-user-ref}. We do not compare against ramp loss \cite{Bottou2006}, as \cite{DingVishy} already attempted it in a similar setting on the majority of data sets we use and were unable to produce any salient results. Not surprisingly, this is an example of the inadequacy of convex optimization methods with respect to non-convex problems. Also, we do not compare against 0-1 loss because it is not margin-enforcing. It is well known that if minimized, 0-1 loss yields the lowest possible training error, but due to the lack of margin enforcement, generalization is bad even when regularization is applied \cite{Vapnik}.

With all methods we perform a standard $10$-fold cross-validation procedure \cite{Dietterich} for locating appropriate values of parameters affecting generalization. Fig.~\ref{fig:error_vs_noise} presents the main results with an emphasis on the consistently superior performance of $q$-loss across all data, especially at high levels of noise. We have verified that often in the high noise cases Tabu search fails to reach the lowest attainable objective value. Therefore we believe we are looking precisely at cases of computationally hard optimizations that fail classically but may be solved successfully by quantum means. We note sigmoid and probit are sometimes close competitors of $q$-loss but other times are the worst performers. This can be explained by their non-convexity, which gives them the potential for robustness, but makes them hard to optimize reliably. However, unlike $q$-loss, we do not know of any AQO-compatible formulations for probit and sigmoid.

\begin{figure}[t]
\centering
\includegraphics[width=8.5cm]{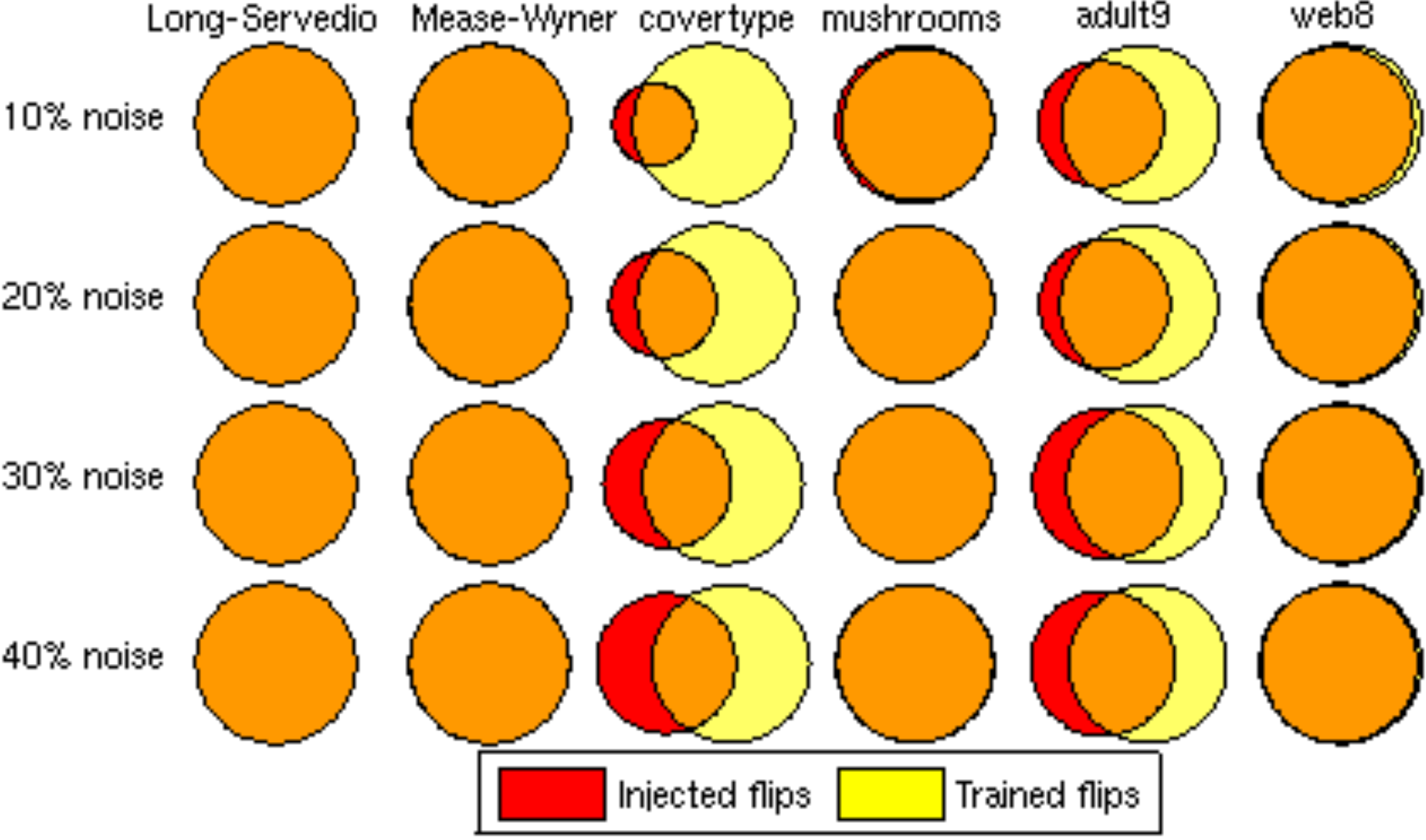}
\caption{Venn diagrams showing overlap between flips injected in the data before training (injected flips) and flips indicated by training with $q$-loss (trained flips). Orange color shows portion of injected flips recovered by $q$-loss.}
\label{fig:venn}
\end{figure}

$q$-loss allows us to identify training examples with possibly incorrect labels as the points with $m \leq q$. We recorded the points whose labels we flipped before training (\emph{injected flips}) and the points that $q$-loss flagged as mislabeled (\emph{trained flips}). Fig.~\ref{fig:venn} summarizes the overlaps between these two sets. The sets of trained flips for covertype and adult9 are expectedly larger due to the large Bayes error of these data sets. In the supplementary material we provide details on cross-validated hyper-parameter values (Sections~\ref{q_t_values} and~\ref{regularization_strength}) and statistical significance tests for the observed error rates (Section~\ref{statistical_significance}).

\section{Conclusion}\label{conclusion}
In this paper we introduced $q$-loss as a robust alternative to convex losses that suffer in the presence of label noise. The QUBO format of the optimization incorporating $q$-loss makes this version of training an ideal candidate for applying emerging commercial AQO technology as the optimization method of choice. Moreover just by using a classical heuristic solver as a stand-in for hardware-based AQO, we were already able to show significant advantages in test error over a rich variety of data sets and across a number of existing convex and non-convex losses. Our focus here was on formulating a robust loss that can be made compatible with the engineering constraints imposed by emerging quantum hardware. Since with other non-convex losses there is no other choice but to resort to often failing convex optimization, $q$-loss stands out with its AQO compliance. This opens up new possibilities for achieving results better than ever seen before. Given such encouraging results, we see great potential for robust classification with $q$-loss under AQO. 

Even though \eqref{q_loss} is a QUBO, future work still needs to address the fact that on large data sets this formulation may result in a number of binary variables that exceeds the available physical qubits. For that reason, options for training via repeated rounds of optimization--e.g. flavors of large neighborhood search---need to be studied. By using suitable graph embedding techniques, we also need to address the fact that not all quadratic interactions between QUBO variables have corresponding connections between qubits on the physical device. Also, future work needs to investigate the asymptotic scaling of the time necessary for optimizing $q$-loss with AQO, similarly to the way that was done for square loss in \cite{Neven09}. Finally, an interesting open question is whether the derivation in Section~\ref{q_loss_section} can be extended to expressing a more general class of functions as QUBOs. 

\section*{Acknowledgments}
We would like to thank Alessandro Bissacco for providing the ``OCR in photos" dataset. We acknowledge Mark Cummins, Edward Farhi, William G. MacReady, James Philbin, and Mani Ranjbar for helpful discussions.



\onecolumn
\appendix
\newpage
\section{Explicit QUBO for training problem}\label{explicit_qubo}
Here we explicitly demonstrate the QUBO problem resulting from training with empirical risk over $q$-loss and Euclidian regularization. To that end we use the variational approximation of $q$-loss and discretize the optimization variables.

Because the discretization itself is cumbersome and uninstructive, in Subsection~\ref{expansion_continuous} we first expand all terms with (mostly) continuous variables and show the general layout of the coefficient matrix $\vc{Q}$ for the QUBO problem
\begin{align}\label{general_qubo}
\min_{\vc{\omega}}\vc{\omega}^T \vc{Q} \vc{\omega} = \min_{\vc{\omega}}\left\{\sum_{\substack{i, j\\i \neq j}}\omega_i \omega_j [Q_{i,j}] + \sum_{i}\omega_i [Q_{i, i}] \right\}\hbox{ ,}
\end{align}
where in our case $\vc{\omega}$ is a concatenation of the binary representations of the discretized $N$ weight variables $w$, the bias $b$, and the $S$ variational parameters $t$. With the notation in \eqref{general_qubo} we adopt the convention of distinguishing problem coefficients $Q_{*,*}$ by placing them inside square brackets. The preceding symbols are always the corresponding variables.

Finally, in Subsection~\ref{binary_variables} we replace the continuous variables $\vc{w}$, $b$, $\vc{t}$ respectively by their discretized versions $\dot{\vc{w}}$, $\dot{b}$, $\dot{\vc{t}}$ according to bit depths $d_w$, $d_b$, $d_t$ and multiplier-offset pairs $(\alpha_w, \beta_w)$, $(\alpha_b, \beta_b)$, $(\alpha_t, \beta_t)$ that determine the intervals in which the discrete variables take values.

\subsection{Expansion with continuous variables}\label{expansion_continuous}
Using the variational approximation for $q$-loss, the empirical risk expands as
\begin{align}\label{expanded_empirical_risk}
\frac{1}{S}\sum_{s=1}^S L_q(m_s) = \frac{1}{S}\sum_{s=1}^S \min_{t_s} \left\{m_s^2  - 2m_s t_s + t_s^2 + \left(1 - q\right)^2\frac{\left(1 - \sign\left(t_s - 1\right)\right)}{2}\right\} \hbox{ .}
\end{align}
Now we expand the individual terms appearing on the right-hand side of \eqref{expanded_empirical_risk}. The goal is to distinguish the coefficients of the various terms in the optimization problem. Hence, in terminal expressions for each term we use the square brackets convention of \eqref{general_qubo}.
\begin{align}
m_s^2 = (\vc{w}^T\vc{x}_s + b)^2 &= \sum_{\substack{i=1\\j=1}}^N w_i w_j[x_{s,i}x_{s,j}] + bb[1] + b\sum_{i=1}^N w_i [2x_{s,i}]\\
-2m_s t_s = -2 y_s(\vc{w}^T\vc{x}_s + b) t_s &= t_s\sum_{i=1}^N w_i [-2y_s x_{s,i}] + b t_s [-2y_s]\\
t_s^2 &= t_s t_s [1]\\
\left(1 - q\right)^2\frac{\left(1 - \sign\left(t_s - 1\right)\right)}{2} = (1 - t_{s,d_t})(1-q)^2 &=  t_{s,d_t}\left[-(1-q)^2\right] + (1-q)^2\label{msb_trick}
\end{align}
The idea behind \eqref{msb_trick} is to use the most significant bit $t_{s,d_t}$ in the binary expansion of $t_s$ as an indicator of $\sign(t_s - 1)$. We give more details on that in Subsection~\ref{binary_variables}.

The above individual expansions when summed over $s$ become:
\begin{align}
\sum_{s=1}^S m_s^2 &= \sum_{\substack{i=1\\j=1}}^N w_i w_j\left[\sum_{s=1}^S x_{s,i}x_{s,j}\right] + bb[S] + b\sum_{i=1}^N w_i\left[2\sum_{s=1}^S x_{s,i}\right]\\
\sum_{s=1}^S -2m_s t_s &= \sum_{i=1}^N\sum_{s=1}^S w_i t_s [-2y_s x_{s,i}] + b \sum_{s=1}^S t_s [-2y_s]\\
\sum_{s=1}^S t_s^2 &= \sum_{s=1}^S t_s t_s[1]\\
\sum_{s=1}^S \left(1 - q\right)^2\frac{\left(1 - \sign\left(t_s - 1\right)\right)}{2} &= \sum_{s=1}^S t_{s,d_t}\left[-(1-q)^2\right] + S(1-q)^2\label{msb_trick_s}
\end{align}

Finally, we can write down the terminally expanded version of the training problem. Note that due to the variational approximation of $q$-loss, we now have a joint optimization problem over the weight variables $\vc{w}$, the bias $b$, and the variational parameters $\vc{t}$.
\begin{align}
&(\vc{w}, b, \vc{t})^{*} = \arg\min_{\vc{w}, b, \vc{t}} \left\{
\sum_{\substack{i=1\\j=1}}^N w_i w_j \underbrace{\left[\frac{1}{S}\sum_{s=1}^S x_{s,i}x_{s,j}\right]}_{\mathcal{A}_{i,j}} + bb\underbrace{[1]}_{\mathcal{B}} + b\sum_{i=1}^N w_i\underbrace{\left[\frac{2}{S}\sum_{s=1}^S x_{s,i}\right]}_{\mathcal{C}_{i}} + \right.\nonumber \\
&\left.+ \sum_{\substack{i=1\\\\}}^N\sum_{s=1}^S w_i t_s\underbrace{\left[\frac{-2y_s x_{s,i}}{S}\right]}_{\mathcal{D}_{i,s}} + b \sum_{s=1}^S t_s \underbrace{\left[\frac{-2y_s}{S}\right]}_{\mathcal{E}_{s}} + \sum_{s=1}^St_s t_s \underbrace{\left[\frac{1}{S}\right]}_{\mathcal{F}_{s}} + \sum_{s=1}^S t_{s,d_t}\underbrace{\left[\frac{-(1-q)^2}{S}\right]}_{\mathcal{G}_{s}} + \sum_{i=1}^N w_i w_i\underbrace{[\lambda]}_{\mathcal{H}_{i}}
\right\}\label{qubo_continuous}
\end{align}

\begin{figure}[h!]
\centering
\includegraphics[width=8.5cm]{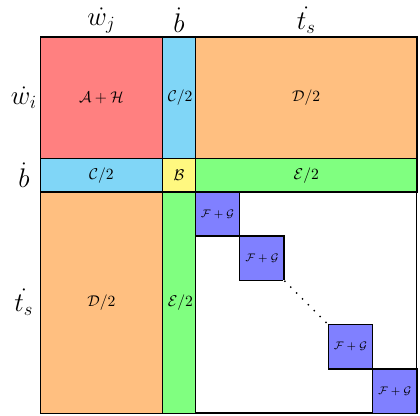}
\caption{Overall layout of the coefficient matrix $\vc{Q}$ implied by \eqref{qubo_continuous} for the QUBO problem $\min_{\vc{\omega}}\vc{\omega}^T \vc{Q} \vc{\omega}$, where $\vc{\omega}$ consists of the concatenation of the binary representations of the discretized variables $\dot{\vc{w}}$, $\dot{b}$, and $\dot{\vc{t}}$. White areas in the matrix correspond to zero coefficients.}
\label{fig:coeff_matrix_diagram}
\end{figure}

In \eqref{qubo_continuous} we dropped the term $S(1-q)^2$ coming from \eqref{msb_trick_s} because it only represents a constant offset. Fig.~\ref{fig:coeff_matrix_diagram} shows the overall layout of the coefficient matrix implied by the coefficient groups $\mathcal{A}$-$\mathcal{H}$ distinguished from \eqref{qubo_continuous}. Note that \eqref{qubo_continuous} is still using the continuous variables (except for the bits $t_{s,d_t}$), but it is clear that after discretizing we can obtain the final QUBO from it.

\subsection{Binary variables}\label{binary_variables}
The final step for obtaining a QUBO is to discretize the continuous variables $\vc{w}$, $b$, $\vc{t}$ via binary expansions of bit-depth $d_w$, $d_b$, $d_t$ respectively. We denote the discrete variables by $\dot{\vc{w}}$, $\dot{b}$, $\dot{\vc{t}}$. We also define multiplier-offset pairs $(\alpha_w, \beta_w)$, $(\alpha_b, \beta_b)$, $(\alpha_t, \beta_t)$ that determine the intervals in which the discrete variables take values.

We apply discretizing transformations by binary variables $w_{*,*}$, $b_{*}$, $t_{*,*}$ and the shorthand function\\$\delta_{\{w,b,t\}}(k) = 2^{k-1}/(2^{d_{\{w,b,t\}}} - 1)$:
\begin{align}
w_i \rightarrow \dot{w}_i &= \alpha_w \sum_{k=1}^{d_w} w_{i,k}\delta_w(k) + \beta_w \hbox { for } i = 1,\ldots,N\\
b \rightarrow \dot{b} &= \alpha_b \sum_{k=1}^{d_b} b_k \delta_b(k) + \beta_b\\
t_s \rightarrow \dot{t}_s &= \alpha_t \sum_{k=1}^{d_t} t_{s,k}\delta_t(k) + \beta_t \hbox { for } s = 1,\ldots,S
\end{align}
The intervals in which the discrete variables take values are:
\begin{align*}
&\dot{w}_i \in [\beta_w; \alpha_w + \beta_w]\\
&\dot{b} \in [\beta_b; \alpha_b + \beta_b]\\
&\dot{t}_s \in [\beta_t; \alpha_t + \beta_t]
\end{align*}
As shown in \eqref{msb_trick} and \eqref{msb_trick_s}, we take the most significant bit $t_{s,d_t}$ of each variable $\dot{t}_s$ as an indicator for $\sign(\dot{t}_s - 1)$. Therefore we need to choose the interval in which the variables $\dot{\vc{t}}$ take values such that the most significant bit of each $\dot{t}_s$ is zero for $\dot{t}_s < 1$ and one otherwise. This leads us to intervals for which the upper half of the representable values are greater than or equal to one, so we set $\beta_{\{w,b,t\}} = 1 - \frac{\alpha_{\{w,b,t\}}}{2}$, which gives the intervals in terms of $\alpha_{\{w,b,t\}} > 0$ only:
\begin{align}
\dot{w}_i &\in \left[1-\frac{\alpha_{w}}{2}; 1+\frac{\alpha_{w}}{2}\right]\\
\dot{b} &\in \left[1-\frac{\alpha_{b}}{2}; 1+\frac{\alpha_{b}}{2}\right]\\
\dot{t}_s &\in \left[1-\frac{\alpha_{t}}{2}; 1+\frac{\alpha_{t}}{2}\right]\label{t_interval}
\end{align}

Now we convert the various terms in \eqref{qubo_continuous} from continuous to binary variables, which gives the coefficients of the final QUBO problem:
\begin{align*}
\sum_{\substack{i=1\\j=1}}^N w_i w_j \left[\mathcal{A}_{i,j}\right] &\rightarrow\\
&\rightarrow \sum_{\substack{i=1\\j=1}}^N \left(\alpha_w \sum_{k=1}^{d_w} w_{i,k}\delta_w(k) + \beta_w\right) \left(\alpha_w \sum_{k'=1}^{d_w} w_{j,k'}\delta_w(k') + \beta_w\right) \left(\mathcal{A}_{i,j}\right) =\\
&=\sum_{\substack{i=1\\j=1}}^N \sum_{\substack{k=1\\k'=1}}^{d_w} w_{i,k}w_{j,k'}\left[\alpha_w^2\delta_w(k)\delta_w(k')\mathcal{A}_{i,j}\right] + \sum_{i=1}^N \sum_{k=1}^{d_w} w_{i,k} \left[2\alpha_w\beta_w\delta_w(k)\sum_{\substack{j=1}}^N\mathcal{A}_{i,j}\right] + \beta_w^2\sum_{\substack{i=1\\j=1}}^N\mathcal{A}_{i,j}\\
bb[\mathcal{B}] &\rightarrow\\
&\rightarrow \left(\alpha_b \sum_{k=1}^{d_b} b_k \delta_b(k) + \beta_b\right)\left(\alpha_b \sum_{k'=1}^{d_b} b_k' \delta_b(k') + \beta_b\right)(\mathcal{B}) =\\
&= \sum_{\substack{k=1\\k'=1}}^{d_b} b_k b_{k'}\left[\alpha_b^2\delta_b(k)\delta_b(k')\mathcal{B}\right] + \sum_{k=1}^{d_b}b_k\left[2\alpha_b\beta_b\delta_b(k)\mathcal{B}\right] + \beta_b^2\mathcal{B}\\
b\sum_{i=1}^N w_i\left[\mathcal{C}_{i}\right] &\rightarrow\\
&\rightarrow \left(\alpha_b \sum_{k=1}^{d_b} b_k \delta_b(k) + \beta_b\right)\sum_{i=1}^N \left(\alpha_w \sum_{k'=1}^{d_w} w_{i,k'}\delta_w(k') + \beta_w\right) \left(\mathcal{C}_{i}\right) =\\
&= \sum_{i=1}^N\sum_{k=1}^{d_b}\sum_{k'=1}^{d_w}b_kw_{i,k'}\left[\alpha_b\alpha_w\delta_b(k)\delta_w(k')\mathcal{C}_{i}\right] + \sum_{i=1}^N\sum_{k'=1}^{d_w}w_{i,k'}\left[\alpha_w\beta_b\delta_w(k')\mathcal{C}_{i}\right] +\\
&+\sum_{k=1}^{d_b}b_k\left[\alpha_b\beta_w\delta_b(k)\sum_{i=1}^N\mathcal{C}_{i}\right] + \beta_b\beta_w\sum_{i=1}^N\mathcal{C}_{i}\\
\sum_{\substack{i=1\\\\}}^N\sum_{s=1}^S w_i t_s\left[\mathcal{D}_{i,s}\right] &\rightarrow\\
& \rightarrow \sum_{\substack{i=1\\\\}}^N\sum_{s=1}^S \left(\alpha_w \sum_{k=1}^{d_w} w_{i,k}\delta_w(k) + \beta_w\right)\left(\alpha_t \sum_{k'=1}^{d_t} t_{s,k'}\delta_t(k') + \beta_t\right)\left(\mathcal{D}_{i,s}\right) =\\
&= \sum_{i=1}^N \sum_{s=1}^S \sum_{k=1}^{d_w}\sum_{k'=1}^{d_t}w_{i,k}t_{s,k'}\left[\alpha_w\alpha_t\delta_w(k)\delta_t(k')\mathcal{D}_{i,s}\right] + \sum_{i=1}^N \sum_{k=1}^{d_w} w_{i,k}\left[\alpha_w\beta_t\delta_w(k)\sum_{s=1}^S\mathcal{D}_{i,s}\right] +\\
&+\sum_{s=1}^S\sum_{k'=1}^{d_t}t_{s,k'}\left[\alpha_t\beta_w\delta_t(k')\sum_{i=1}^N\mathcal{D}_{i,s}\right] + \beta_w\beta_t\sum_{i=1}^N \sum_{s=1}^S\mathcal{D}_{i,s}\\
b \sum_{s=1}^S t_s \left[\mathcal{E}_{s}\right] &\rightarrow \\
&\rightarrow \left(\alpha_b \sum_{k=1}^{d_b} b_k \delta_b(k) + \beta_b\right) \sum_{s=1}^S \left(\alpha_t \sum_{k'=1}^{d_t} t_{s,k'}\delta_t(k') + \beta_t\right) \left(\mathcal{E}_{s}\right)=\\
&=\sum_{s=1}^S\sum_{k=1}^{d_b}\sum_{k'=1}^{d_t}  b_k t_{s,k'}\left[\alpha_b\alpha_t\delta_b(k)\delta_t(k')\mathcal{E}_{s}\right] + \sum_{k=1}^{d_b}b_k\left[\alpha_b\beta_t\delta_b(k)\sum_{s=1}^S\mathcal{E}_{s}\right] +\\
&+ \sum_{s=1}^S\sum_{k'=1}^{d_t}t_{s,k'}\left[\alpha_t\beta_b\delta_t(k')\mathcal{E}_{s}\right] + \beta_b\beta_t\sum_{s=1}^S\mathcal{E}_{s}
\end{align*} 
\begin{align*}
\sum_{s=1}^St_s t_s \left[\mathcal{F}_{s}\right] &\rightarrow\\
&\rightarrow \sum_{s=1}^S \left(\alpha_t \sum_{k=1}^{d_t} t_{s,k}\delta_t(k) + \beta_t\right) \left(\alpha_t \sum_{k'=1}^{d_t} t_{s,k'}\delta_t(k') + \beta_t\right) \left(\mathcal{F}_{s}\right) =\\
&=\sum_{s=1}^S\sum_{\substack{k=1\\k'=1}}^{d_t} t_{s,k}t_{s,k'}\left[\alpha_t^2\delta_t(k)\delta_t(k')\mathcal{F}_{s}\right] + \sum_{s=1}^S\sum_{k=1}^{d_t}t_{s,k}\left[2\alpha_t\beta_t\delta_t(k)\mathcal{F}_{s}\right] + \beta_t^2\sum_{s=1}^S\mathcal{F}_{s}\\
\sum_{i=1}^N w_i w_i[\mathcal{H}_{i}] &\rightarrow \\
&\rightarrow \sum_{i=1}^N \left(\alpha_w \sum_{k=1}^{d_w} w_{i,k}\delta_w(k) + \beta_w\right) \left(\alpha_w \sum_{k'=1}^{d_w} w_{i,k'}\delta_w(k') + \beta_w\right)(\mathcal{H}_{i})=\\
&= \sum_{i=1}^N \sum_{\substack{k=1\\k'=1}}^{d_w} w_{i,k}w_{i,k'}\left[\alpha_w^2\delta_w(k)\delta_w(k')\mathcal{H}_{i}\right] + \sum_{i=1}^N \sum_{k=1}^{d_w} w_{i,k}\left[2\alpha_w\beta_w\delta_w(k)\mathcal{H}_{i}\right] + \beta_w^2\sum_{i=1}^N\mathcal{H}_{i}
\end{align*}

\section{Data summary}\label{data_summary}

\begin{table}[H]
\caption{Summary of data sets}
\centering
\begin{tabular}{ccccccc}
\toprule
Name & Dims & \#Examples & Density ($\%$) & Baseline error ($\%$) & $d_{w}$\\
\midrule
\rowcolor[gray]{0.9}
Long-Sevedio & 21 & 2000 & 100.00 & 50.00 & 2\\
Mease-Wyner & 20 & 2000 & 100.00 & 49.80 & 2\\
\rowcolor[gray]{0.9}
covertype & 54 & 581012 & 22.20 & 36.46 & 4\\
mushrooms & 112 & 8124 & 18.75 & 48.20 & 4\\
\rowcolor[gray]{0.9}
adult9 & 123 & 48842 & 11.30 & 23.93 & 4\\
web8 & 300 & 59245 & 4.20 & 2.92 & 4\\
\bottomrule
\end{tabular}\label{table:data_summary}
\end{table}

\section{$q$ values for $q$-loss and $t$ values for $t$-logistic}\label{q_t_values}

\begin{table}[H]
\caption {Approximate lower bounds for $q$ in $q$-loss computed according to Section 5. In each case the $q$ values offered to cross-validation are taken as the 10 equally spaced values between the bound and 0 (both inclusive).}
\centering
\begin{tabular}{cccccc}
\toprule
\multirow{2}{*}{Data set name} & \multicolumn{5}{c}{Label noise ($\%$)} \\
\cmidrule(r){2-6}
& 0 & 10 & 20 & 30 & 40 \\
\midrule
\rowcolor[gray]{0.9}
Long-Servedio  & -1000 & -3.486401 & -2.172365 & -1.590225 & -1.243201\\
Mease-Wyner & -25.726124 & -3.333979 & -2.084934 & -1.524450 & -1.188680\\
\rowcolor[gray]{0.9}
covertype & -1.133948 & -0.979198 & -0.853870 & -0.749685 & -0.661297\\
mushrooms & -69.710678 & -3.400772 & -2.114833 & -1.544074 & -1.203589\\
\rowcolor[gray]{0.9}
adult9 & -1.552383 & -1.201211 & -0.963926 & -0.789849 & -0.655127\\
web8 & -8.901475 & -2.081794 & -1.233931 & -0.839671 & -0.600122\\
\bottomrule
\end{tabular}\label{table:q_bounds}
\end{table}

\begin{table}[H]
\begin{minipage}[b]{0.5\linewidth}
\caption {$q$ values for $q$-loss picked by cross-validation}
\centering
\begin{tabular}{cccccc}
\toprule
\multirow{2}{*}{Data set name} & \multicolumn{5}{c}{Label noise ($\%$)} \\
\cmidrule(r){2-6}
& 0 & 10 & 20 & 30 & 40 \\
\midrule
\rowcolor[gray]{0.9}
Long-Servedio  & 0 & -0.39 & -0.24 & -0.71 & -0.55\\
Mease-Wyner & 0 & -2.96 & -1.62 & -1.36 & 0\\
\rowcolor[gray]{0.9}
covertype & -0.63 & -0.54 & -0.38 & -0.5 & -0.51\\
mushrooms & 0 & -0.76 & -0.47 & -0.17 & -0.13\\
\rowcolor[gray]{0.9}
adult9 & -0.86 & -0.53 & -0.43 & -0.53 & -0.07\\
web8 & -0.99 & -0.46 & -0.41 & -0.19 & 0\\
\bottomrule
\end{tabular}\label{table:q_value}
\end{minipage}
\hspace{0.5cm}
\begin{minipage}[b]{0.5\linewidth}
\caption {$t$ values for t-logistic picked by cross-validation}
\centering
\begin{tabular}{cccccc}
\toprule
\multirow{2}{*}{Data set name} & \multicolumn{5}{c}{Label noise ($\%$)} \\
\cmidrule(r){2-6}
& 0 & 10 & 20 & 30 & 40 \\
\midrule
\rowcolor[gray]{0.9}
Long-Servedio  & 1.1 & 1.1 & 1.5 & 1.9 & 2.0\\
Mease-Wyner & 1.1 & 1.6 & 2.0 & 1.9 & 1.9\\
\rowcolor[gray]{0.9}
covertype & 1.1 & 1.2 & 1.5 & 1.3 & 1.2\\
mushrooms & 1.1 & 1.1 & 1.2 & 1.1 & 1.1\\
\rowcolor[gray]{0.9}
adult9 & 1.2 & 1.3 & 2.0 & 1.2 & 2.0\\
web8 & 1.1 & 1.1 & 1.1 & 1.2 & 1.1\\
\bottomrule
\end{tabular}\label{table:t_value}
\end{minipage}
\end{table}

\section{Regularization strength}\label{regularization_strength}
\begin{table}[H]
\caption {$\lambda$ values offered to cross-validation ($C$ for liblinear is $1/\lambda$)}
\centering
\begin{tabular}{c}
\toprule
$\lambda$ \\
\midrule
\rowcolor[gray]{0.9}
2.000090\\
0.398965\\
\rowcolor[gray]{0.9}
0.079583\\
0.015875\\
\rowcolor[gray]{0.9}
0.003167\\
0.000632\\
\rowcolor[gray]{0.9}
0.000126\\
0.000025\\
\rowcolor[gray]{0.9}
0.000005\\
0.000001\\
\bottomrule
\end{tabular}\label{table:lambda_values}
\end{table}

\begin{table}[H]
\caption {$C$ values for liblinear picked by cross-validation}
\centering
\begin{tabular}{cccccc}
\toprule
\multirow{2}{*}{Data set name} & \multicolumn{5}{c}{Label noise ($\%$)} \\
\cmidrule(r){2-6}
& 0 & 10 & 20 & 30 & 40 \\
\midrule
\rowcolor[gray]{0.9}
Long-Servedio  & 0.499978 & 2.506486 & 0.499978 & 0.499978 & 0.499978\\
Mease-Wyner & 40000.000000 & 0.499978 & 315.756236 & 12.565498 & 62.992126\\
\rowcolor[gray]{0.9}
covertype & 0.499978 & 2.506486 & 62.992126 & 1000000.000000 & 12.565498\\
mushrooms & 2.506486 & 12.565498 & 0.499978 & 0.499978 & 0.499978\\
\rowcolor[gray]{0.9}
adult9 & 0.499978 & 62.992126 & 0.499978 & 0.499978 & 0.499978\\
web8 & 315.756236 & 0.499978 & 12.565498 & 12.565498 & 0.499978\\
\bottomrule
\end{tabular}\label{table:c_liblinear}
\end{table}

\begin{table}[H]
\caption {$\lambda$ values picked by cross-validation for $0\%$ label noise}
\centering
\begin{tabular}{cccccccc}
\toprule
\multirow{2}{*}{Data set name} & \multicolumn{7}{c}{Method} \\
\cmidrule(r){2-8}
& $q$ & logistic & square & t-logistic & sigmoid & probit & smooth hinge\\
\midrule
\rowcolor[gray]{0.9}
Long-Servedio & 0.015875 & 0.003167 & 0.079583 & 0.003167 & 0.000632 & 0.003167 & 0.015875\\
Mease-Wyner & 0.000126 & 0.000001 & 0.000025 & 0.000001 & 0.000025 & 0.003167 & 0.000001\\
\rowcolor[gray]{0.9}
covertype & 0.000025 & 0.000025 & 0.000025 & 0.000001 & 2.000090 & 2.000090 & 0.000001\\
mushrooms & 0.000025 & 0.000001 & 0.000025 & 0.000126 & 0.000632 & 0.015875 & 0.000632\\
\rowcolor[gray]{0.9}
adult9 & 0.003167 & 0.000001 & 0.000632 & 0.000025 & 0.003167 & 0.003167 & 0.000126\\
web8 & 0.000632 & 0.000001 & 0.000005 & 0.000005 & 0.000126 & 0.000632 & 0.000001\\
\bottomrule
\end{tabular}\label{table:lambda_noise0}
\end{table}

\begin{table}[H]

\caption {$\lambda$ values picked by cross-validation for $10\%$ label noise}
\centering
\begin{tabular}{cccccccc}
\toprule
\multirow{2}{*}{Data set name} & \multicolumn{7}{c}{Method} \\
\cmidrule(r){2-8}
& $q$ & logistic & square & t-logistic & sigmoid & probit & smooth hinge\\
\midrule
\rowcolor[gray]{0.9}
Long-Servedio & 0.015875 & 0.000005 & 2.000090 & 0.000126 & 0.000632 & 0.003167 & 0.003167\\
Mease-Wyner & 0.000126 & 0.000005 & 0.000632 & 0.000001 & 0.000632 & 0.003167 & 0.000005\\
\rowcolor[gray]{0.9}
covertype & 0.000001 & 0.000025 & 0.000632 & 0.000001 & 0.000632 & 0.003167 & 0.000126\\
mushrooms & 0.003167 & 0.000005 & 0.000001 & 0.000001 & 0.000632 & 0.003167 & 0.000005\\
\rowcolor[gray]{0.9}
adult9 & 0.015875 & 0.000632 & 0.003167 & 0.000632 & 0.003167 & 0.003167 & 0.000126\\
web8 & 0.000632 & 0.000005 & 0.000126 & 0.000001 & 0.000126 & 0.000632 & 0.000005\\
\bottomrule
\end{tabular}\label{table:lambda_noise10}
\end{table}

\begin{table}[H]

\caption {$\lambda$ values picked by cross-validation for $20\%$ label noise}
\centering
\begin{tabular}{cccccccc}
\toprule
\multirow{2}{*}{Data set name} & \multicolumn{7}{c}{Method} \\
\cmidrule(r){2-8}
& $q$ & logistic & square & t-logistic & sigmoid & probit & smooth hinge\\
\midrule
\rowcolor[gray]{0.9}
Long-Servedio & 0.000126 & 2.000090 & 2.000090 & 0.000025 & 0.000632 & 0.003167 & 2.000090\\
Mease-Wyner & 0.000126 & 0.000025 & 0.000005 & 0.000001 & 0.003167 & 0.003167 & 0.000126\\
\rowcolor[gray]{0.9}
covertype & 0.000001 & 0.000001 & 0.000126 & 0.000001 & 0.000025 & 0.000632 & 0.000126\\
mushrooms & 0.003167 & 0.000126 & 0.000632 & 0.000126 & 0.000632 & 0.003167 & 0.000025\\
\rowcolor[gray]{0.9}
adult9 & 0.015875 & 0.079583 & 0.079583 & 0.003167 & 0.000632 & 0.003167 & 0.003167\\
web8 & 0.000632 & 0.000001 & 0.000001 & 0.000005 & 2.000090 & 0.000632 & 0.000126\\
\bottomrule
\end{tabular}\label{table:lambda_noise20}
\end{table}

\begin{table}[H]

\caption {$\lambda$ values picked by cross-validation for $30\%$ label noise}
\centering
\begin{tabular}{cccccccc}
\toprule
\multirow{2}{*}{Data set name} & \multicolumn{7}{c}{Method} \\
\cmidrule(r){2-8}
& $q$ & logistic & square & t-logistic & sigmoid & probit & smooth hinge\\
\midrule
\rowcolor[gray]{0.9}
Long-Servedio & 0.003167 & 2.000090 & 2.000090 & 0.000001 & 0.000126 & 0.003167 & 2.000090\\
Mease-Wyner & 0.000126 & 0.000005 & 0.000001 & 0.000001 & 0.003167 & 0.003167 & 0.000005\\
\rowcolor[gray]{0.9}
covertype & 0.000025 & 0.000001 & 0.000126 & 0.000001 & 0.000632 & 0.003167 & 0.000025\\
mushrooms & 0.003167 & 0.000632 & 0.003167 & 0.000632 & 0.003167 & 0.003167 & 0.000632\\
\rowcolor[gray]{0.9}
adult9 & 0.003167 & 2.000090 & 0.003167 & 2.000090 & 0.000126 & 0.000632 & 2.000090\\
web8 & 0.000632 & 0.000126 & 0.000001 & 0.000632 & 0.000632 & 0.003167 & 0.000126\\
\bottomrule
\end{tabular}\label{table:lambda_noise30}
\end{table}

\begin{table}[H]
\caption {$\lambda$ values picked by cross-validation for $40\%$ label noise}
\centering
\begin{tabular}{cccccccc}
\toprule
\multirow{2}{*}{Data set name} & \multicolumn{7}{c}{Method} \\
\cmidrule(r){2-8}
& $q$ & logistic & square & t-logistic & sigmoid & probit & smooth hinge\\
\midrule
\rowcolor[gray]{0.9}
Long-Servedio & 0.003167 & 2.000090 & 2.000090 & 0.000001 & 0.000126 & 0.000632 & 2.000090\\
Mease-Wyner & 0.000126 & 0.000001 & 0.000005 & 0.000001 & 2.000090 & 0.003167 & 0.000001\\
\rowcolor[gray]{0.9}
covertype & 0.000001 & 0.000001 & 0.000001 & 0.000001 & 2.000090 & 2.000090 & 0.000001\\
mushrooms & 0.003167 & 0.000126 & 0.000632 & 0.000001 & 0.003167 & 0.015875 & 0.003167\\
\rowcolor[gray]{0.9}
adult9 & 0.000025 & 0.000126 & 0.000126 & 0.000001 & 0.000632 & 0.003167 & 0.079583\\
web8 & 0.000632 & 0.015875 & 0.079583 & 0.015875 & 0.000632 & 0.000632 & 0.000632\\
\bottomrule
\end{tabular}\label{table:lambda_noise40}
\end{table}

\section{Statistical significance}\label{statistical_significance}
\begin{table}[H]
\caption {Paired $t$-test for statistical significance of the test error difference yielded by $q$-loss and an estimated closest competitor. The closest competitor is manually chosen on a per-data-set basis from the set of convex losses. We exclude the other non-covex losses (t-logistic, sigmoid, and probit) from this comparison as we do not know what their performance would be if they could be realiably solved to optimality. We reject the null hypothesis at $\alpha = 5\%$ significance level. 'Y' means that the difference is significant and 'N' means the difference is not significant.}
\centering
\begin{tabular}{ccccccc}
\toprule
\multirow{2}{*}{Data set name} & \multirow{2}{*}{Compared losses} & \multicolumn{5}{c}{Label noise ($\%$)} \\
\cmidrule(r){3-7}
& & 0 & 10 & 20 & 30 & 40 \\
\midrule
\rowcolor[gray]{0.9}
Long-Servedio  & smoothed hinge vs $q$-loss & N & N & Y & Y & Y\\
Mease-Wyner & smoothed hinge vs $q$-loss & Y & Y & Y & Y & Y\\
\rowcolor[gray]{0.9}
covertype & liblinear vs $q$-loss  & Y & Y & Y & Y & Y\\
mushrooms & smoothed hinge vs $q$-loss & N & N & N & N & Y\\
\rowcolor[gray]{0.9}
adult9 & smoothed hinge vs $q$-loss & Y & Y & Y & Y & Y\\
web8 & smoothed hinge vs $q$-loss & Y & Y & Y & Y & Y\\
\bottomrule
\end{tabular}\label{table:significance}
\end{table}

\end{document}